\documentclass[journal]{IEEEtran}

\usepackage{amsfonts}
\usepackage{cite}
\usepackage{graphicx}
\usepackage{amssymb}
\usepackage{subfigure}
\usepackage{array}
\usepackage{color}
\usepackage{amsmath}
\usepackage{algorithmic}
\usepackage{algorithm}
\usepackage{cases}
\usepackage{bm}
\usepackage{mathrsfs}
\usepackage{epsfig}
\usepackage{psfrag}
\usepackage{url}
\usepackage{setspace}

\usepackage{etoolbox}
\makeatletter
\patchcmd{\@makecaption}
  {\scshape}
  {}
  {}
  {}
\makeatletter
\patchcmd{\@makecaption}
  {\\}
  {.\ }
  {}
  {}
\makeatother

\newtheorem{theorem}{Theorem}

\newtheorem{corollary}{Corollary}

\newtheorem{Prob}{Problem}
\newtheorem{remark}{Remark}

\newcounter{TempEqCnt}

\IEEEoverridecommandlockouts

\begin{document}

\title{Analysis and Optimization of an Intelligent Reflecting Surface-assisted System with Interference}


\author{Yuhang Jia, {\em Graduate Student Member, IEEE}, Chencheng Ye, and Ying Cui, {\em Member, IEEE}
\thanks{Manuscript received January  29, 2020; revised May 04, 2020; accepted August 17, 2020. This work was supported in part by the National Key R$\&$D Program of China under Grant 2018YFB1801102, and Natural Science Foundation of Shanghai under Grant 20ZR1425300.
The paper has been presented in part at the IEEE ICC 2020 \cite{Analysis}. The associate editor coordinating the review of this paper and approving it for publication was Luiz DaSilva. \textit{(Corresponding author: Ying Cui.)}

Yuhang Jia, Chencheng Ye and Y. Cui are with the Department of Electronic Engineering,
Shanghai Jiao Tong University, Shanghai 200240, China (e-mail:cuiying@sjtu.edu.cn).

}}

\maketitle

\begin{abstract}
In this paper, we study an intelligent reflecting surface (IRS)-assisted system where a multi-antenna base station (BS) serves a single-antenna user with the help of a multi-element IRS in the presence of interference generated by a multi-antenna BS serving its own single-antenna user. The signal and interference links via the IRS are modeled with Rician fading. To reduce phase adjustment cost, we adopt quasi-static phase shift design where the phase shifts do not change with the instantaneous channel state information (CSI). We investigate two cases of CSI at the BSs, namely, the instantaneous CSI case and the statistical CSI case, and apply Maximum Ratio Transmission (MRT) based on the complete CSI and the CSI of the Line-of-sight (LoS) components, respectively. Different costs on channel estimation and beamforming adjustment are incurred in the two CSI cases. First, we obtain a tractable expression of the average rate in the instantaneous CSI case and a tractable expression of the ergodic rate in the statistical CSI case. We also provide sufficient conditions for the average rate in the instantaneous CSI case to surpass the ergodic rate in the statistical CSI case, at any phase shifts. Then, we maximize the average rate and ergodic rate, both with respect to the phase shifts, leading to two non-convex optimization problems. For each problem, we obtain a globally optimal solution under certain system parameters, and propose an iterative algorithm based on parallel coordinate descent (PCD) to obtain a stationary point under arbitrary system parameters. Next, in each CSI case, we provide sufficient conditions under which the optimal quasi-static phase shift design is beneficial, compared to the system without IRS. Finally, we numerically verify the analytical results and demonstrate notable gains of the proposal solutions over existing ones. To the best of our knowledge, this is the first work that considers optimal quasi-static phase shift design for an IRS-assisted system in the presence of interference.
\end{abstract}


%
\IEEEpeerreviewmaketitle

\section{Introduction}
With the deployment of the fifth-generation (5G) wireless network, the urgent requirement on network capacity is gradually being achieved. But the increasingly demanding requirement on energy efficiency remains unaddressed.
Recently, intelligent reflecting surface (IRS), consisting of nearly passive,
low-cost, reflecting elements with reconfigurable parameters, is  envisioned to serve as a promising solution for improving spectrum and energy efficiency \cite{IEEEexample:8910627,IEEEexample:8796365}.
Experimental results have also demonstrated significant gains of IRS-assisted systems over systems without IRSs \cite{IEEEexample:7510962,IEEEexample:8485924}.

In \cite{IEEEexample:zheng2019intelligent,IEEEexample:8683663
,IEEEexample:you2019intelligent,IEEEexample:8879620,IEEEexample:nadeem2019large,IEEEexample:yang2019intelligent,IEEEexample:guo2019weighted,8941080,IEEEexample:8855810
,IEEEexample:yu2019enabling,IEEEexample:8741198,
IEEEexample:8811733,Guo,IEEEexample:zhang2019analysis,IEEEexample:8746155}, the authors consider IRS-assisted systems where one multi-antenna base station (BS) serves one or multiple users with the help of one multi-element IRS
\cite{IEEEexample:zheng2019intelligent,IEEEexample:8683663
,IEEEexample:you2019intelligent,IEEEexample:8879620,IEEEexample:yang2019intelligent,IEEEexample:guo2019weighted,8941080,IEEEexample:8855810
,IEEEexample:yu2019enabling,IEEEexample:8741198,
IEEEexample:8811733,Guo,IEEEexample:8746155}, \cite{IEEEexample:nadeem2019large},
or multiple multi-element IRSs \cite{IEEEexample:zhang2019analysis}. In \cite{IEEEexample:zheng2019intelligent,IEEEexample:8683663
,IEEEexample:you2019intelligent,IEEEexample:8879620}, the authors assume block fading channels and investigate the estimation of instantaneous channel states. For instance, \cite{IEEEexample:8683663
,IEEEexample:zheng2019intelligent,IEEEexample:you2019intelligent} estimate the channel state of the indirect link via each element of the IRS by switching on the IRS elements one by one; \cite{IEEEexample:8879620} focuses on cascaded channel estimation of the indirect links via all elements of the IRS, based on carefully pre-designed phase shifts for the IRS elements.  In \cite{IEEEexample:yang2019intelligent,IEEEexample:guo2019weighted,8941080,IEEEexample:8855810
,IEEEexample:yu2019enabling,IEEEexample:8741198,
IEEEexample:8811733}, the authors investigate the joint optimization of the beamformer at the BS and the phase shifts at the IRS to maximally improve system performance. In \cite{IEEEexample:jiang2019over,IEEEexample:8743496,
IEEEexample:8742603,IEEEexample:cui2019secure,
IEEEexample:yan2019passive}, various other IRS-assisted systems are studied. For example, in \cite{IEEEexample:jiang2019over}, the authors propose to boost the performance of over-the-air computation with the help of a multi-element IRS. In \cite{IEEEexample:8743496,IEEEexample:8742603,
IEEEexample:cui2019secure}, the authors consider a system where a multi-antenna BS servers multiple single-antenna legitimate users in the presence of eavesdroppers, with the help of a multi-element IRS. In \cite{IEEEexample:yan2019passive}, the authors consider a system where a multi-element IRS assists the primary communication from a single-antenna user to a multi-antenna BS and sends information to the BS at the same time.

According to whether the phase shifts are adaptive to instantaneous channel state information (CSI) or not, these works \cite{IEEEexample:8855810,IEEEexample:guo2019weighted,8941080,IEEEexample:nadeem2019large
,IEEEexample:yu2019enabling,IEEEexample:yang2019intelligent,IEEEexample:8746155,IEEEexample:8743496,IEEEexample:8742603,
IEEEexample:cui2019secure,IEEEexample:zhang2019analysis,
IEEEexample:jiang2019over,IEEEexample:yan2019passive,
IEEEexample:8811733,IEEEexample:8741198,Guo} can be classified into two categories.
In one category \cite{IEEEexample:8855810,IEEEexample:guo2019weighted,8941080
,IEEEexample:yu2019enabling,IEEEexample:yang2019intelligent,IEEEexample:8743496,IEEEexample:8742603,
IEEEexample:cui2019secure,
IEEEexample:jiang2019over,
IEEEexample:8811733,IEEEexample:8741198,IEEEexample:yan2019passive}, phase shifts are adjusted based on instantaneous CSI which is assumed to be known. For instance, in \cite{IEEEexample:8855810,IEEEexample:guo2019weighted,8941080
,IEEEexample:yu2019enabling,IEEEexample:yang2019intelligent,IEEEexample:8743496,IEEEexample:8742603,
IEEEexample:cui2019secure,
IEEEexample:jiang2019over,
IEEEexample:8811733,IEEEexample:8741198,IEEEexample:yan2019passive}, the authors consider the maximization of the sum rate \cite{IEEEexample:yang2019intelligent,IEEEexample:8743496,IEEEexample:8742603,
IEEEexample:cui2019secure,IEEEexample:yan2019passive},  weighted sum rate \cite{IEEEexample:guo2019weighted,8941080} or  energy efficiency \cite{IEEEexample:8855810,IEEEexample:yu2019enabling,IEEEexample:8741198}, and the minimization of the transmit power \cite{IEEEexample:jiang2019over,IEEEexample:8811733}.
The aforementioned
optimization problems are all non-convex. The authors propose iterative algorithms to obtain locally optimal solutions or nearly optimal solutions of the non-convex problems in \cite{IEEEexample:8855810,IEEEexample:guo2019weighted,8941080
,IEEEexample:yu2019enabling,IEEEexample:yang2019intelligent,
IEEEexample:8811733,IEEEexample:8741198,IEEEexample:8743496,
IEEEexample:8742603,IEEEexample:cui2019secure,
IEEEexample:jiang2019over,IEEEexample:yan2019passive}.

In the other category \cite{IEEEexample:nadeem2019large,IEEEexample:8746155,IEEEexample:zhang2019analysis,Guo}, phase shifts are determined by statistics of CSI and do not change with instantaneous CSI.  In \cite{Guo}, \cite{IEEEexample:zhang2019analysis}, the authors consider slowly varying Non-line-of-sight (NLoS) components, and minimize the outage probability. The optimization problems are non-convex. In contrast, in \cite{IEEEexample:8746155}, \cite{IEEEexample:nadeem2019large}, the authors consider fast varying NLoS components, and maximize the ergodic rate \cite{IEEEexample:8746155} or the minimum ergodic rate. By analyzing problem structures, closed-form optimal phase shifts are obtained for the non-convex problems in \cite{Guo,IEEEexample:8746155,IEEEexample:zhang2019analysis} or an approximate problem of the non-convex problem in \cite{IEEEexample:nadeem2019large}. Compared with instantaneous
CSI-adaptive phase shift designs in the first category, quasi-static phase shift designs in the second one have lower implementation costs, owing to less frequent phase adjustment.

Note that all the aforementioned works \cite{Guo,IEEEexample:8855810,IEEEexample:guo2019weighted,8941080,IEEEexample:nadeem2019large
,IEEEexample:yu2019enabling,IEEEexample:yang2019intelligent,
IEEEexample:8811733,IEEEexample:8741198,IEEEexample:8743496,
IEEEexample:8742603,IEEEexample:cui2019secure,
IEEEexample:jiang2019over,IEEEexample:yan2019passive,IEEEexample:8746155,IEEEexample:zhang2019analysis}  ignore interference from other transmitters, when investigating IRS-assisted communications. However, in practical wireless networks, interference usually has a severe impact, especially in dense networks or for cell-edge users. It is thus critical to take into account the role of interference in designing IRS-assisted systems. In \cite{IEEEexample:pan2019intelligent}, the authors optimize the instantaneous CSI-adaptive phase shift design and beamforming at the signal BS to maximize the weighted sum rate of an IRS-assisted system in the presence of  an interference BS. In \cite{pan}, the authors optimize the instantaneous CSI-adaptive phase shift design and beamformers at all BSs to maximize the weighted sum rate in an IRS-assisted multi-cell network with inter-cell interference. As the instantaneous CSI-adaptive designs in \cite{IEEEexample:pan2019intelligent,pan} have higher phase adjustment costs, it is highly desirable to obtain cost-efficient quasi-static phase shift design for IRS-assisted systems with interference. Furthermore, it is also important to characterize the gain derived from IRS in systems with interference.

In this article, we shall shed some light on the aforementioned issues. We consider an IRS-assisted system where a multi-antenna BS serves a single-antenna user with the help of a multi-element IRS, in the presence of interference generated by a multi-antenna BS serving its own single-antenna user. The antennas at the two BSs and the reflecting elements at the IRS are arranged in uniform rectangular arrays (URAs). The signal and interference links via the IRS are modeled with Rician fading, while the links between the BSs and the users are modeled with Rayleigh fading. As in \cite{IEEEexample:8746155}, \cite{IEEEexample:nadeem2019large}, we assume that the line-of-sight (LoS) components do not change but the NLoS components vary fast during the considered time duration. To reduce phase adjustment cost, we adopt quasi-static phase shift design, where the phase shifts do not change with instantaneous CSI, but only adapt to  CSI statistics. We investigate two cases of CSI at the BSs, namely, the instantaneous CSI case and the statistical CSI case, where different costs on channel estimation and beamforming adjustment are inccured. In the two CSI cases, we apply Maximum Ratio Transmission (MRT) based on the complete CSI (i.e., the CSI of both the LoS and NLoS components) and the CSI of the NLoS components, respectively. In this paper, we focus on the analysis and optimization of the average rate in the instantaneous CSI case and the ergodic rate in the statistical CSI case for the IRS-assisted transmission in the presence of interference. The theoretical results offer important insights for designing practical IRS-assisted systems. The main contributions of the article are summarized as follows.
\begin{itemize}
\item First, we obtain a tractable expression of the average rate in the instantaneous CSI case and a tractable expression of the ergodic rate in the statistical CSI case. We show that under certain conditions, the average rate in the instantaneous CSI case is greater than the ergodic rate in the statistical CSI case, at any phase shifts, demonstrating the value of the CSI of NLoS components in performance improvement via beamforming.

\item  Then, we optimize the phase shifts to maximize the average rate in the instantaneous CSI case and the ergodic rate in the statistical CSI case, respectively, leading to two non-convex optimization problems. Under certain system parameters, we obtain a globally optimal solution of each non-convex problem. Under arbitrary system parameters, we propose an iterative algorithm based on parallel coordinate descent (PCD), to obtain a stationary point of each non-convex problem. The proposed PCD algorithm is particularly suitable for systems with large-scale IRS and multi-core processors which support parallel computing, compared with the state-of-the-art algorithms, i.e., the block coordinate descent (BCD) algorithm and the minorization maximization (MM) algorithm \cite{IEEEexample:yu2019enabling,IEEEexample:pan2019intelligent,pan}.  Furthermore, we characterize the average rate degradation and ergodic rate degradation caused by the quantization error for the phase shifts.

\item Next, in each CSI case, we provide sufficient conditions under which the optimal quasi-static phase shift design (with the minimum phase adjustment cost for the IRS-assisted system) is beneficial in the presence of interference, compared to a counterpart system without IRS.

\item Finally, by numerical results, we verify analytical results and demonstrate notable gains of the proposed solutions over existing schemes. We also reveal the specific value of  the PCD algorithm for large-scale IRS.
\end{itemize}



\section{System Model}
\begin{figure}[t]
\begin{center}
 \includegraphics[width=5.5cm]{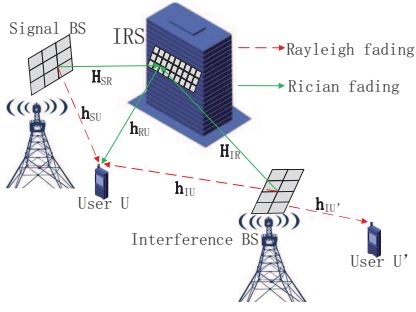}
  \end{center}
     \caption{\small{System Model.}}
\label{fig:system model}
\end{figure}

As shown in Fig. \ref{fig:system model},
one single-antenna user, say user $U$, is served by a BS with the help of an IRS. The BS is referred to as the signal BS of user $U$ or BS $S$. The IRS is installed on the wall of a high-rise building. Another BS serving its own single-antenna user, say user $U'$, causes interference to user $U$, and hence is referred to as the interference BS of user $U$ or BS $I$. The signal BS and the IRS are far from user $U'$. The signal BS and interference BS are equipped with URAs of $M_S\times N_S$ antennas and $M_I\times N_I$ antennas, respectively. Assume $M_SN_S>1$ and $M_IN_I>1$. The IRS is equipped with a URA of $M_R\times N_R$ reflector elements. Without loss of generality, we assume $M_c\leq N_c$, where $c=S, I$. For notation simplicity, define $\mathcal M_c \triangleq \{1,2,...,M_c\}$ and $\mathcal N_c \triangleq \{1,2,...,N_c\}$, where $c=S,I,R$. Suppose that the two users do not move during a certain time period. In this paper, we wound like to investigate how the signal BS serves user $U$ with the help of the IRS in the presence of interference.

As scattering is often rich near the ground, we adopt the Rayleigh model for  the channels between the BSs and the users. Let $\mathbf{h}^H_{SU} \in \mathbb{C}^{ 1 \times M_SN_S }$, $\mathbf{h}^H_{IU} \in \mathbb{C}^{1 \times M_IN_I}$ and $\mathbf{h}^H_{IU'} \in \mathbb{C}^{1 \times M_IN_I}$ denote the channel vectors for the channel between the signal BS and user $U$, the channel between the interference BS and user $U$, and the channel between the interference BS and user $U'$, respectively. Specifically,
\begin{align*}
\mathbf{h}^H_{cU}=&\sqrt{{\alpha_{cU}}}\tilde{\mathbf{h}}^H_{cU},
\quad \mathbf{h}^H_{IU'}=\sqrt{{\alpha_{IU'}}}\tilde{\mathbf{h}}^H_{IU'},
\end{align*}where $c=S,I$, $\alpha_{cU}$, $\alpha_{IU'}>0$ represent the distance-dependent path losses, and the elements of $\tilde{\mathbf{h}}^H_{cU}$, $\tilde{\mathbf{h}}^H_{IU'}$ are independent and identically distributed according to $C\mathcal N(0,1)$.

As scattering is much weaker far from the ground, we adopt the Rician fading model for the channels between the BSs and the IRS and the channel between user $U$ and the IRS. Let $\mathbf{H}_{SR} \in \mathbb{C}^{M_RN_R \times M_SN_S}$, $\mathbf{H}_{IR} \in \mathbb{C}^{M_RN_R \times M_IN_I}$ and $\mathbf{h}^{H}_{RU} \in \mathbb{C}^{1 \times M_RN_R}$ denote the channel matrices for the channel between the signal BS and the IRS, the channel between the interference BS and the IRS and the channel between the IRS and user $U$, respectively. Specifically,
\begin{align*}
\mathbf{H}_{cR}=&\sqrt{\alpha_{cR}} \left(\sqrt{\frac{K_{cR}}{K_{cR}+1}} \bar{\mathbf{H}}_{cR} + \sqrt{\frac{1}{K_{cR}+1}} \tilde{\mathbf{H}}_{cR}\right),\\
\mathbf{h}^{H}_{RU} = & \sqrt{\alpha_{RU}} \left(\sqrt{\frac{K_{RU}}{K_{RU}+1}} \bar{\mathbf{h}}^{H}_{RU} + \sqrt{\frac{1}{K_{RU}+1}} \tilde{\mathbf{h}}^{H}_{RU}\right),
\end{align*}
where $c=S, I$, $\alpha_{cR}$, $\alpha_{RU}>0$ represent the distance-dependent path losses, $K_{cR}$, $K_{RU}\geq 0$ denote the Rician factors,\footnote{If $K_{SR}=0$, $K_{IR}=0$, or $K_{RU}=0$, the corresponding Rician fading reduces down to Rayleigh fading. If $K_{SR} \rightarrow \infty$, $K_{IR} \rightarrow \infty$, or $K_{RU} \rightarrow \infty$, only the LoS component exists.} $\tilde{\mathbf{H}}_{cR} \in \mathbb{C}^{M_RN_R \times M_cN_c}$ and $\tilde{\mathbf{h}}^{H}_{RU} \in \mathbb{C}^{1 \times M_RN_R }$ represent the normalized NLoS components, with elements independently and identically distributed according to $C\mathcal N(0,1)$, and $\bar{\mathbf{H}}_{cR} \in \mathbb{C}^{M_RN_R \times M_cN_c}$
and $\bar{\mathbf{h}}^{H}_{RU}\in \mathbb{C}^{1\times M_RN_R}$ represent the deterministic normalized LoS components, with unit-modulus elements. Note that $\bar{\mathbf{H}}_{cR}$ and $\bar{\mathbf{h}}^{H}_{RU}$ do not change during the considered time period, as the location of user $U$ is assumed to be invariant.

Let $\lambda$ and $d$ $(\leq \frac{\lambda}{2})$ denote the wavelength of transmission signals and the distance between adjacent elements or antennas  in each row and each column of the URAs. Define:
\begin{align}
& f(\theta^{(h)},\theta^{(v)},m,n)\nonumber \\ \triangleq  & 2\pi\frac{d}{\lambda}\sin\theta^{(v)}((m-1)\cos\theta^{(h)}+(n-1)\sin\theta^{(h)}), \label{eq:f} \\
 &\mathbf{A}_{m,n}(\theta^{(h)},\theta^{(v)},M,N)
 \nonumber\\ \triangleq & \left(e^{jf(\theta^{(h)},\theta^{(v)},m,n)}\right)_{m=1,...,M,n=1,...,N}  ,
\\
 &\mathbf{a}(\theta^{(h)},\theta^{(v)},M,N)
 \nonumber \\
 \triangleq & \text{rvec}\left( \mathbf{A}_{m,n}(\theta^{(h)},\theta^{(v)},M,N) \right),
 \end{align}
Here, $\mathbf{A}_{m,n}(\theta^{(h)},\theta^{(v)},M,N) \in \mathbb{C}^{M \times N}$, $\mathbf{a}(\theta^{(h)},\theta^{(v)},M,N) \in \mathbb{C}^{1 \times MN}$, and rvec($\cdot$) denotes the row vectorization of a matrix.
Then, $\bar{\mathbf{H}}_{cR}$
and $\bar{\mathbf{h}}^{H}_{RU}$ are modeled as \cite{IEEEexample:van2004optimum}:
\begin{align*}
\bar{\mathbf{H}}_{cR} = & \mathbf{a}^H(\delta^{(h)}_{cR},\delta^{(v)}_{cR},M_R,N_R) \mathbf{a}(\varphi^{(h)}_{cR},\varphi^{(v)}_{cR},M_c,N_c),\\
\bar{\mathbf{h}}_{RU}^H = & \mathbf{a}(\varphi^{(h)}_{RU},\varphi^{(v)}_{RU},M_R,N_R),
\end{align*}
where $c=S,I$. Here, $\delta^{(h)}_{cR}$ $\left(\delta^{(v)}_{cR}\right)$ represents the azimuth (elevation) angle between the direction of a row (column) of the URA at the IRS and the projection of the signal from BS $c$ to the IRS on the plane of the URA at the IRS; $\varphi^{(h)}_{cR}$ $\left(\varphi^{(v)}_{cR}\right)$ represents the azimuth (elevation) angle between the direction of a row (column) of the URA at BS $c$ and the projection of the signal from BS $c$ to the IRS on the plane of the URA at BS $c$; $\varphi^{(h)}_{RU}$ $\left(\varphi^{(v)}_{RU}\right)$ represents the azimuth (elevation) angle between the direction of a row (column) of the URA at the IRS and the projection of the signal from the IRS to user $U$ on the plane of the URA at the IRS.

To reduce phase adjustment cost, we consider quasi-static phase shift design where the phase shifts do not change with the NLoS components, which vary fast. Let $\boldsymbol\phi \triangleq \left(\phi_{m,n}\right)_{m \in \mathcal M_R, n \in \mathcal N_R}\in \mathbb{C}^{M_R \times N_R}$ represent the constant phase shifts of the IRS with $\phi_{m,n}$ being the phase shift of the $(m,n)$-th element of the IRS, where
\begin{align}
\phi_{m,n} \in [0,2\pi),\quad m \in \mathcal M_R, n \in \mathcal N_R.\label{eq:phi}
\end{align}For convenience, define
\setcounter{TempEqCnt}{\value{equation}}
\setcounter{equation}{8}
\begin{figure*}[t]
\begin{align}
\gamma^{(instant)}(\boldsymbol\phi) = & \frac{{P_S}\left\lvert\left\lvert  \mathbf{h}^H_{RU}\Phi(\boldsymbol\phi) \mathbf{H}_{SR} + \mathbf{h}^H_{SU}  \right\rvert\right\rvert_2^2}{ {P_I}\mathbb{E} \left[ {\left\lvert (\mathbf{h}^H_{RU}\Phi(\boldsymbol\phi) \mathbf{H}_{IR} + \mathbf{h}^H_{IU})  \frac{\mathbf{h}_{IU'}}{\left\lvert\left\lvert \mathbf{h}_{IU'} \right\rvert\right\rvert_2} \right\rvert}^2 \right]+ {\sigma}^2} \label{eq:sinr}
\end{align}
\hrulefill
\vspace{-1mm}
\end{figure*}
\setcounter{equation}{\value{TempEqCnt}}
$\Phi(\boldsymbol\phi) \triangleq \text{diag} \left(\text{rvec}\left(\left(e^{j\phi_{m,n}}\right)_{m \in \mathcal M_R, n \in \mathcal N_R}\right)\right) \in \mathbb{C}^{M_RN_R\times M_RN_R}$,
where diag($\cdot$) denotes a square diagonal matrix with the elements of a vector on the main diagonal.
 We focus on the IRS-assisted transmission from the signal BS to user $U$ in the presence of the interference BS. The channel of the indirect link between BS $c$ and user $U$ via the IRS is given by $\mathbf{h}^H_{RU}\Phi(\boldsymbol\phi) \mathbf{H}_{cR}$, and hence, the equivalent channel between BS $c$ and user $U$ is given by $\mathbf{h}^H_{RU}\Phi(\boldsymbol\phi) \mathbf{H}_{cR} + \mathbf{h} ^H_{cU}$, where $c=S,I$. We consider linear beamforming at the signal BS and interference BS for serving user $U$ and user $U'$, respectively. Let $\mathbf{w}_S \in \mathbb{C}^{M_SN_S \times 1}$ and $\mathbf{w}_I \in \mathbb{C}^{M_IN_I \times 1}$ denote the corresponding normalized beamforming vectors, where ${\left\lvert\left\lvert \mathbf{w}_S \right\rvert\right\rvert}^2_2=1$ and ${\left\lvert\left\lvert \mathbf{w}_I\right\rvert\right\rvert}^2_2=1$.
Thus, the signal received at user $U$ is expressed as:
\begin{align}
\begin{split}
Y \triangleq & \sqrt{P_S}(\mathbf{h}^H_{RU}\Phi(\boldsymbol\phi) \mathbf{H}_{SR} + \mathbf{h} ^H_{SU}) \mathbf{w}_S X_S
\\ &+ \sqrt{P_I}\left(\mathbf{h}^H_{RU}\Phi(\boldsymbol\phi) \mathbf{H}_{IR} + \mathbf{h}^H_{IU}\right) \mathbf{w}_I X_I + Z,
\end{split}
\end{align}
where $P_S$ and $P_I$ are the transmit powers of the signal BS and interference BS, respectively, $X_S$ and $X_I$ are the information symbols for user $U$ and user $U'$, respectively, with $\mathbb{E}\left[{\lvert X_S\rvert}^2\right] = 1$ and $\mathbb{E}\left[{\lvert X_I \rvert}^2\right] = 1$, and $Z \sim C\mathcal{N}(0,\sigma^2)$ is the additive white gaussian noise (AWGN). Assume that user $U$ knows $(\mathbf{h}^H_{RU}\Phi(\boldsymbol\phi) \mathbf{H}_{SR} + \mathbf{h} ^H_{SU}) \mathbf{w}_S$, but does not know $\left(\mathbf{h}^H_{RU}\Phi(\boldsymbol\phi) \mathbf{H}_{IR} + \mathbf{h}^H_{IU}\right) \mathbf{w}_I$.
 In the following, we consider two cases, namely the instantaneous CSI case and the statistical CSI case, where different costs on channel estimation and beamforming adjustment are incurred and different system performances can be achieved.
\subsection{Instantaneous CSI Case}
In this part, assume that the CSI of the equivalent channel between the signal BS and user $U$, i.e., $\mathbf{h}^H_{RU}\Phi(\boldsymbol\phi) \mathbf{H}_{SR} + \mathbf{h}^H_{SU}$, is known at the signal BS, and the CSI of the channel between the interference BS and user $U'$, i.e., $\mathbf{h}_{IU'}$, is known at the interference BS. Note that for any given $\boldsymbol\phi$,\footnote{Later, we shall see that $\boldsymbol\phi$ can be determined based on some known system parameters.} $\mathbf{h}^H_{RU}\Phi(\boldsymbol\phi) \mathbf{H}_{SR} + \mathbf{h}^H_{SU}$ can be directly estimated by the signal BS via a pilot sent by user $U$, and $\mathbf{h}_{IU'}$ can be estimated by the interference BS via a pilot sent by user $U'$\cite{IEEEexample:zheng2019intelligent,IEEEexample:8683663
,IEEEexample:you2019intelligent,IEEEexample:8879620}. This case is referred to as the instantaneous CSI case.

In the instantaneous CSI case, to enhance the signals received at user $U$ and user $U'$, respectively, we consider the instantaneous CSI-adaptive MRT at the signal BS and interference BS, respectively:\footnote{It is obvious that $\mathbf{w}^{(instant)}_{S}$ in \eqref{eq:bv s} is optimal for the maximization of $\frac{{P_S}{\left\lvert\left(  \mathbf{h}^H_{RU}\Phi(\boldsymbol\phi) \mathbf{H}_{SR} + \mathbf{h}^H_{SU}\right)\mathbf{w}_{S}\right\rvert}^2}{ {P_I}\mathbb{E} \left[ \left\lvert \left(\mathbf{h}^H_{RU}\Phi(\boldsymbol\phi) \mathbf{H}_{IR} + \mathbf{h}^H_{IU}\right)\mathbf{w}_{I} \right\rvert^2 \right]+ {\sigma}^2}$, with respect to $\mathbf{w}_{S}$ under $\left\lvert\left\lvert\mathbf{w}_{S}\right\rvert\right\rvert_2^2=1$, for any $\boldsymbol\phi$ and $\mathbf{w}_{I}$. Thus, $\mathbf{w}^{(instant)}_{S}$ is optimal for the average rate maximization.}
\begin{align}
\mathbf{w}^{(instant)}_{S} = &\ \frac{{\left(\mathbf{h}^H_{RU}\Phi(\boldsymbol\phi) \mathbf{H}_{SR} + \mathbf{h}^H_{SU}\right)}^H}{\left\lvert\left\lvert \mathbf{h}^H_{RU}\Phi(\boldsymbol\phi) \mathbf{H}_{SR} + \mathbf{h}^H_{SU} \right\rvert\right\rvert_2}\label{eq:bv s},\\
\mathbf{w}^{(instant)}_{I} = &\ \frac{\mathbf{h}_{IU'}}{\left\lvert \left\lvert \mathbf{h}_{IU'} \right\rvert\right \rvert_2}\label{eq:bv i}.
\end{align}
Here, $\mathbf{w}^{(instant)}_{c} \in \mathbb{C}^{M_cN_c \times 1}$, $c=S,I$. In the instantaneous CSI case, the achievable rate\footnote{Note that $\left(\mathbf{h}^H_{RU}\Phi(\boldsymbol\phi) \mathbf{H}_{IR} + \mathbf{h}^H_{IU}\right) \mathbf{w}_I$ is not known at user $U$. By treating $\left(\mathbf{h}^H_{RU}\Phi(\boldsymbol\phi) \mathbf{H}_{IR} + \mathbf{h}^H_{IU}\right) \mathbf{w}_IX_I \sim C\mathcal{N}\left(0,\mathbb{E}\left[\left\lvert\left(\mathbf{h}^H_{RU}\Phi(\boldsymbol\phi) \mathbf{H}_{IR} + \mathbf{h}^H_{IU}\right) \mathbf{w}_I\right\rvert^2\right]\right)$, which corresponds to the worst-case noise, $\log_2\left(1+\gamma^{(instant)}(\boldsymbol\phi)\right)$ can be achieved.} is $\log_2\left(1+\gamma^{(instant)}(\boldsymbol\phi)\right),$ where the signal to interference plus noise ratio (SINR) at user $U$, i.e., $\gamma^{(instant)}(\boldsymbol\phi)$, is given by \eqref{eq:sinr}, as shown at the top of the page.
Therefore, in the instantaneous CSI case, the average rate for the IRS-assisted transmission with interference is given by:
\begin{align}
C^{(instant)}(\boldsymbol\phi) \triangleq \mathbb{E}\left[\log_2 \left(1 +  \gamma^{(instant)}(\boldsymbol\phi)\right)\right],\label{eq:bv20}
\end{align}
where $\gamma^{(instant)}(\boldsymbol\phi)$ is given by \eqref{eq:sinr} and the expectation is with respect to the random NLoS components.
\begin{remark}[Instantaneous CSI-Adaptive MRT without Interference]
When there is no interference BS, i.e., $P_I=0$, $C^{(instant)}(\boldsymbol\phi)$ in \eqref{eq:bv20} reduces to the average rate for the IRS-assisted transmission without interference, in the instantaneous CSI case. Its analysis and optimization under the uniform linear array (ULA) model for the multi-antenna BS (i.e., $M_S=1$ or $N_S=1$) and multi-element IRS (i.e., $M_R=1$ or $N_R=1$) have been investigated in \cite{IEEEexample:8746155}.
\end{remark}
\setcounter{TempEqCnt}{\value{equation}}
\setcounter{equation}{12}
\begin{figure*}[t]
\begin{align}
 \gamma^{(statistic)}(\boldsymbol\phi)= & \frac{{P_S}{\left\lvert   \left(\mathbf{h}^H_{RU}\Phi(\boldsymbol\phi) \mathbf{H}_{SR} + \mathbf{h}^H_{SU}\right) \frac{{\left(\bar{\mathbf{h}}^H_{RU}\Phi(\boldsymbol\phi) \bar{\mathbf{H}}_{SR}\right)}^H}{\left\lvert \left\lvert \bar{\mathbf{h}}^H_{RU}\Phi(\boldsymbol\phi) \bar{\mathbf{H}}_{SR}\right  \rvert \right\rvert_2} \right\rvert}^2}{ {P_I}\mathbb{E} \left[ {\left\lvert \left(\mathbf{h}^H_{RU}\Phi(\boldsymbol\phi) \mathbf{H}_{IR} + \mathbf{h}^H_{IU}\right)\frac{1}{\sqrt{M_IN_I}} \boldsymbol{1}_{M_IN_I} \right\rvert}^2 \right]+ {\sigma}^2}\label{eq:sinr_sta}
\end{align}
\hrulefill
\vspace{-1mm}
\end{figure*}
\setcounter{TempEqCnt}{\value{equation}}
\setcounter{equation}{21}
\begin{figure*}
\begin{align}
A_{SRU,NLoS}^{(Q)} \triangleq &
\begin{cases}
P_SM_SN_S\alpha_{SR}\alpha_{RU}M_RN_R(1-\tau_{SRU}), &Q=instant\\
P_SM_SN_S\alpha_{SR}\alpha_{RU}M_RN_R\left(1-\tau_{SRU}
-\frac{M_SN_S-1}{M_SN_S(K_{SR}+1)}\right), &Q =statistic
\end{cases} \label{eq:A2} \\
A_{IRU,NLoS}^{(Q)} \triangleq &
\begin{cases}
P_I\alpha_{IR}\alpha_{RU}M_RN_R(1-\tau_{IRU}), & Q=instant \\
P_I\alpha_{IR}\alpha_{RU}M_RN_R\left(1-\tau_{IRU}+
\frac{\tau_{IRU}\left(y_{IR}-M_IN_I\right)}{M_IN_IK_{RU}}\right), & Q=statistic
\end{cases} \label{eq:A5}
\end{align}
\hrulefill
\vspace{-1mm}
\end{figure*}
\setcounter{equation}{\value{TempEqCnt}}
\setcounter{equation}{\value{TempEqCnt}}\subsection{Statistical CSI Case}
In this part, assume that only the CSI of the LoS components  $\bar{\mathbf{h}}_{RU}$, $\bar{\mathbf{H}}_{SR}$ are known at the signal BS, and no channel knowledge is known at the interference BS (recall that the channel between the interference BS and user $U'$ is modeled as Rayleigh fading). Note that $\delta^{(h)}_{SR},\delta^{(v)}_{SR}$, $\varphi^{(h)}_{SR},\varphi^{(v)}_{SR}$ depend only on the placement of the URAs at the signal BS and the IRS as well as the locations of them;   $\delta^{(h)}_{{RU}},\delta^{(v)}_{{RU}}$ depend only on the placement of the URA at the IRS and the locations of the IRS and user $U$. Thus, $\bar{\mathbf{h}}_{RU}, \bar{\mathbf{H}}_{SR}$ can be easily determined. This case is called the statistical CSI case.

In the statistical CSI case, to enhance the signal received at user $U$, we consider statistical CSI-adaptive MRT at the signal BS:\footnote{In Appendix A, we show that $\mathbf{w}^{(statistic)}_{S}$ in \eqref{eq:bv13} is optimal for the maximization of $\frac{\mathbb{E}\left[\left\lvert(\mathbf{h}^H_{RU}\Phi(\boldsymbol\phi) \mathbf{H}_{SR} + \mathbf{h} ^H_{SU}) \mathbf{w}_S\right\rvert^2\right]}{\mathbb{E}\left[\left\lvert\left(\mathbf{h}^H_{RU}\Phi(\boldsymbol\phi) \mathbf{H}_{IR} + \mathbf{h}^H_{IU}\right) \mathbf{w}_I^{(statistic)}\right\rvert^2\right]+\sigma^2}$ with respect to $\mathbf{w}_S$ under $\left\lvert\left\lvert\mathbf{w}_S\right\rvert\right\rvert_2^2=1$, for any $\boldsymbol\phi$. Thus, $\mathbf{w}_S^{(statistic)}$ is approximately optimal for the ergodic rate maximization.}
\setcounter{equation}{9}
\begin{align}
\mathbf{w}^{(statistic)}_{S} =\ \frac{{\left(\bar{\mathbf{h}}^H_{RU}\Phi(\boldsymbol\phi) \bar{\mathbf{H}}_{SR}\right)}^H}{\left\lvert\left \lvert \bar{\mathbf{h}}^H_{RU}\Phi(\boldsymbol\phi) \bar{\mathbf{H}}_{SR} \right \rvert \right\rvert_2}.\label{eq:bv13}
\end{align}
As no channel knowledge is available at the interference BS, we choose:\footnote{In the statistical CSI case, any $\mathbf{w}_{I}$ with $\left\lvert\left\lvert\mathbf{w}_{I}\right\rvert\right\rvert_2^2=1$ achieves the same ergodic rate for user $U'$.}
\begin{align}
\mathbf{w}^{(statistic)}_{I} = \frac{1}{\sqrt{M_IN_I}} \boldsymbol{1}_{M_IN_I}. \label{eq:bv4}
\end{align}
Therefore, in the statistical CSI case, coding over a large number of channel coherence time intervals, we can achieve the ergodic rate for the IRS-adaptive transmission with interference:
\begin{align}
C^{(statistic)}(\boldsymbol\phi) \triangleq \mathbb{E}\left[\log_2\left(1 +   \gamma^{(statistic)}(\boldsymbol\phi)\right)\right],\label{eq:bv24}
\end{align}
where
the SINR at user $U$, i.e., $\gamma^{(statistic)}(\boldsymbol\phi)$, is given by \eqref{eq:sinr_sta}, as shown at the top of the next page.
Here, $\boldsymbol{1}_n$ represents the $n$-dimensional unity column vector.
\begin{remark}[Statistical CSI-Adaptive MRT without Interference]
When there is no interference BS, i.e., $P_I=0$, $C^{(statistic)}(\boldsymbol\phi)$ in \eqref{eq:bv24} reduces to the ergodic rate for the IRS-assisted transmission without interference in the statistical CSI case. Note that its analysis or optimization under the ULA model has not yet been considered.  
\end{remark}
\section{Rate Analysis}
In this section, we analyze the average rate in the instantaneous CSI case and the ergodic rate in the statistical CSI case for the IRS-assisted system in the presence of interference.
Define $\tau_{cRU} \triangleq \frac{K_{cR}K_{RU}}{(K_{cR}+1)(K_{RU}+1)}$, $\theta_{cRU,m,n}\triangleq f\left(\varphi_{RU}^{(h)},\varphi_{RU}^{(v)},m,n\right)-
f\left(\delta_{cR}^{(h)},\delta_{cR}^{(v)},m,n\right)$, $\theta_{IR,m,n} \triangleq f\left(\varphi_{IR}^{(h)},\varphi_{IR}^{(v)},m,n\right), $ $m \in \mathcal M_R, n \in \mathcal N_R$, and
\setcounter{equation}{13}
\begin{align}
y_{cRU}(\boldsymbol\phi) \triangleq  & {\left\lvert \sum_{m=1}^{M_R}\sum_{n=1}^{N_R} e^{j\theta_{cRU,m,n} +j\phi_{m,n}}\right\rvert}^2,\label{eq:over gamma}\\
y_{IR} \triangleq & {\left\lvert \sum_{m=1}^{M_I}\sum_{n=1}^{N_I} e^{j\theta_{IR,m,n}}\right\rvert}^2,\label{eq:over gamma1}
\end{align}
where $y_{cRU}(\boldsymbol\phi) \in [0,M_R^2N_R^2]$, $y_{IR} \in [0,M_I^2N_I^2]$ and $f(\cdot)$ is given by \eqref{eq:f}. Note that $\tau_{cRU}$ increases with $K_{cR}$ and $K_{RU}$. In addition, note that $f\left(\varphi_{RU}^{(h)},\varphi_{RU}^{(v)},m,n\right)$ $\left(f\left(\varphi_{IR}^{(h)},\varphi_{IR}^{(v)},m,n\right)\right)$ represents the difference of the phase change over the LoS component between the $(m,n)$-th element of the IRS (the $(m,n)$-th antenna of the interference BS) and user $U$ (the IRS) and the phase change over the LoS component between the ($1,1$)-th element of the IRS (the (1,1)-th antenna of the interference BS) and user $U$ (the IRS); $f\left(\delta_{cR}^{(h)},\delta_{cR}^{(v)},m,n\right)$ represents the difference of the phase change over the LoS component between BS $c$ and the ($m,n$)-th element of the IRS and the phase change over the LoS component between BS $c$ and the ($1,1$)-th element of the IRS. Finally, note that  $\left\lvert\left\lvert{\bar{\mathbf{h}}}^H_{RU}\Phi(\boldsymbol\phi) \bar{\mathbf{H}}_{cR}\right\rvert\right\rvert_2^2=M_cN_cy_{cRU}(\boldsymbol\phi)$, i.e., $M_cN_cy_{cRU}(\boldsymbol\phi)$ represents the sum channel power of the LoS components of the indirect link between BS $c$ and user $U$ via the IRS. Define:
\begin{align}
A_{SRU,LoS}\triangleq &
P_SM_SN_S\alpha_{SR}\alpha_{RU}\tau_{SRU}, \label{eq:A1}\\
A_{SU}^{(Q)}\triangleq &
\begin{cases}
P_SM_SN_S\alpha_{SU}, &Q=instant,\\
P_S\alpha_{SU}, &Q=statistic,
\end{cases} \label{eq:A3}\\
A_{IRU,LoS}^{(Q)}\triangleq &
\begin{cases}
P_I\alpha_{IR}\alpha_{RU}\tau_{IRU}, &Q=instant, \\
P_I\alpha_{IR}\alpha_{RU}\tau_{IRU}\frac{y_{IR}}
{M_IN_I}, & Q=statistic,
\end{cases} \label{eq:A4} \\
A_{IU}\triangleq & P_I\alpha_{IU} +\sigma^2. \label{eq:A6}
\end{align}
The expressions of $C_{ub}^{(instant)}(\boldsymbol\phi)$ and $C_{ub}^{(statistic)}(\boldsymbol\phi)$ are not tractable. As in \cite{IEEEexample:8746155,IEEEexample:pan2019intelligent,IEEEexample:zheng2019intelligent,IEEEexample:cover2012elements}, using Jensen's inequality,  we can obtain their analytical upper bounds.
\begin{theorem}[Upper Bound of Average or Ergodic Rate]\label{lem:ergodic Case 1 with reflector}For $Q=instant$, $statistic$,
\begin{align}
 C^{(Q)}(\boldsymbol\phi) \leq   \log_2\left(1+ \gamma^{(Q)}_{ub}(\boldsymbol\phi)\right)
 \triangleq  C^{(Q)}_{ub}(\boldsymbol\phi),\label{eq:cub}
\end{align}
where
\begin{align}
\gamma^{(Q)}_{ub}(\boldsymbol\phi) \triangleq
 \frac{A_{SRU,LoS} y_{SRU}(\boldsymbol\phi) + A_{SRU,NLoS}^{(Q)}+A_{SU}^{(Q)} }{A_{IRU,LoS}^{(Q)}y_{IRU}(\boldsymbol\phi) + A_{IRU,NLoS}^{(Q)}+A_{IU}}. \label{eq:ub gamma}
\end{align}
Here, $A_{SRU,NLoS}^{(Q)}$ and $A_{IRU,NLoS}^{(Q)}$ are given by \eqref{eq:A2} and \eqref{eq:A5}, as shown at the top of the page.
\end{theorem}
\begin{IEEEproof}
Please refer to Appendix B.
\end{IEEEproof}

Note that when $P_I=0$, implying $A_{IRU,LoS}^{(Q)}=0$ and $A_{IRU,NLoS}^{(Q)}=0$, $\gamma^{(Q)}_{ub}(\boldsymbol\phi)$ becomes:
\setcounter{equation}{23}
\begin{align}
\gamma_{ub}^{(Q)}(\boldsymbol\phi)=
\frac{A_{SRU,LoS} y_{SRU}(\boldsymbol\phi) + A_{SRU,NLoS}^{(Q)}+A_{SU}^{(Q)}}
{\sigma^2}.\label{eq:withoutinterferenceBS}
\end{align}
Without the interference BS (i.e., $P_I=0$) and with ULAs at the signal BS and IRS (i.e., $M_I=1$ or $N_I=1$ and $M_R=1$ or $N_R=1$), Theorem \ref{lem:ergodic Case 1 with reflector} for $Q=instant$ reduces to Theorem 1 in \cite{IEEEexample:8746155}. Later in Section \ref{sec:simulation}, we shall show that $C_{ub}^{(Q)}(\boldsymbol\phi)$ is a good approximation of $C^{(Q)}(\boldsymbol\phi)$, and  can facilitate the evaluation and optimization for it.

From Theorem \ref{lem:ergodic Case 1 with reflector}, we can draw the following conclusions. For all $\boldsymbol\phi$ and $Q=instant$, $statistic$, $C_{ub}^{(Q)}(\boldsymbol\phi)$ increases with $P_S$, $M_S$, $N_S$, $\alpha_{SR}$ and $\alpha_{SU}$, and decreases with $P_I$, $\alpha_{IR}$, $\alpha_{IU}$ and $\sigma^2$; $C_{ub}^{(Q)}\left(\boldsymbol\phi\right)$ increases with $\gamma_{ub}^{(Q)}(\boldsymbol\phi)$. Thus, we can compare $C_{ub}^{(instant)}\left(\boldsymbol\phi\right)$ and $C_{ub}^{(statistic)}\left(\boldsymbol\phi\right)$ by comparing $\gamma_{ub}^{(instant)}\left(\boldsymbol\phi\right)$ and $\gamma_{ub}^{(statistic)}\left(\boldsymbol\phi\right)$, and maximize $C_{ub}^{(Q)}\left(\boldsymbol\phi\right)$ by maximizing $\gamma^{(Q)}_{ub}\left(\boldsymbol\phi\right)$. Furthermore, by Theorem \ref{lem:ergodic Case 1 with reflector}, we have the following results.
\begin{corollary}\label{cor:respective}
(i) $A_{SRU,NLoS}^{(instant)} > A_{SRU,NLoS}^{(statistic)}$ and
$A_{SU}^{(instant)} > A_{SU}^{(statistic)}$.
(ii) If $P_I > 0$ and $y_{IR}>M_IN_I$, $A_{IRU,LoS}^{(instant)}< A_{IRU,LoS}^{(statistic)}$ and
$A_{IRU,NLoS}^{(instant)} < A_{IRU,NLoS}^{(statistic)}$.
\end{corollary}

Corollary~\ref{cor:respective} (i) implies that the received signal power at user $U$ in the instantaneous CSI case always surpasses that in the statistical CSI case, at any phase shifts. Corollary~\ref{cor:respective} (ii) implies that in the presence of interference, if $y_{IR} >M_IN_I$, the received interference power at user $U$ in the instantaneous CSI case is weaker than that in the statistical CSI case, at any phase shifts. Note that $y_{IR}$ given in \eqref{eq:over gamma1} is a function of $\varphi_{IR}^{(h)}$ and $\varphi_{IR}^{(v)}$, which depend only on the placement of the URA at the interference BS and the locations of the interference BS and the IRS. Corollary~\ref{cor:respective} indicates the value of CSI of the NLoS components in improving the receive SINR at user $U$.
\begin{corollary}\label{cor:inssta}
(i) If $P_I<\varepsilon$ for some $\varepsilon >0$, $\gamma_{ub}^{(instant)}(\boldsymbol\phi)>\gamma_{ub}^{(statistic)}(\boldsymbol\phi)$, for all $\boldsymbol\phi$.
(ii) If $y_{IR}> M_IN_I$, $\gamma_{ub}^{(instant)}(\boldsymbol\phi)>\gamma_{ub}^{(statistic)}(\boldsymbol\phi)$, for all $\boldsymbol\phi$.
\end{corollary}

Corollary~\ref{cor:inssta} (i) means that in the presence of weak interference, the average rate in the instantaneous CSI case is greater than the ergodic rate in the statistical CSI case, at any phase shifts. Corollary~\ref{cor:inssta} (ii) means that if the placement of the URA at the interference BS and the locations of the interference BS and IRS satisfy certain condition, the average rate in the instantaneous CSI case is greater than the ergodic rate in the statistical CSI case, at any phase shifts. Corollary~\ref{cor:inssta} reveals the advantage of CSI of the NLoS components in improving the receive SINR at user $U$.\footnote{Note that $\gamma_{ub}^{(instant)}(\boldsymbol\phi)>\gamma_{ub}^{(statistic)}(\boldsymbol\phi)$ does not always hold, as the interference powers in the two cases are different.}
\section{Rate Optimization}
In this section, we maximize the average rate in the instantaneous CSI case and the ergodic rate in the statistical CSI case for the IRS-assisted system in the presence of interference. Specifically, we would like to maximize the upper bound $C_{ub}^{(Q)}(\boldsymbol\phi)$ of $C^{(Q)}(\boldsymbol\phi)$, or equivalently maximize $\gamma_{ub}^{(Q)}\left(\boldsymbol\phi\right)$ by optimizing the phase shifts $\boldsymbol\phi$ subject to the constraints in \eqref{eq:phi}.
\begin{Prob}[Average or Ergodic Rate Maximization]\label{prob:eq}For $Q=instant$ or $statistic$,
\begin{align*}
\gamma_{ub}^{(Q)*} \triangleq \max_{\boldsymbol\phi} & \  \gamma_{ub}^{(Q)}(\boldsymbol\phi)\\
 s.t.&\ \eqref{eq:phi},
\end{align*}
where $\gamma_{ub}^{(Q)}(\boldsymbol\phi)$ is given by~\eqref{eq:ub gamma}.
Let $\boldsymbol\phi^{(Q)*}$ denote an optimal solution.
\end{Prob}

For $Q=instant$ or $statistic$, an optimal solution depends on the LoS components and the distributions of the NLoS components. In general, $\boldsymbol\phi^{(instant)*}$ and $\boldsymbol\phi^{(statistic)*}$ are different, as different beamformers are applied in the two CSI cases. Note that Problem~\ref{prob:eq} is a challenging non-convex problem. In the following, we tackle Problem~\ref{prob:eq} in some special cases (with certain system parameters) and the general case (with arbitrary system parameters), respectively. We also characterize the impact of the number of quantization bits for the optimal phase shifts on rate degradation.

\subsection{Globally Optimal Solutions in Special Cases}
Define $\Lambda(x) \triangleq  x-2\pi\left\lfloor\frac{x}{2\pi}\right\rfloor, x \in \mathbb{R}
$, and $\eta^{(Q)}\triangleq A_{SRU,LoS}\left(A_{IRU,NLoS}^{(Q)}+A_{IU}\right)- A_{IRU,LoS}^{(Q)}\left(A_{SRU,NLoS}^{(Q)}+A_{SU}^{(Q)}\right)
,Q=instant \text{ or }statistic$.
Note that $\frac{\lvert{x-\Lambda(x)}\rvert}{2\pi} \in \mathbb{N}$ and $\Lambda(x) \in [0,2\pi)$, as $\frac{\Lambda(x)}{2\pi}=\frac{x}{2\pi}-\left\lfloor\frac{x}{2\pi}\right\rfloor \in [0,1)$ for all  $x \in \mathbb{R}$. That is, $\Lambda(\cdot)$ can be used to provide phase shifts $\boldsymbol\phi$ satisfying \eqref{eq:phi}. By the triangle inequality and by analyzing structural properties of Problem \ref{prob:eq}, we obtain globally optimal solutions in four special cases:
\begin{itemize}
\item Special Case (i): $M_R=N_R=1$;
\item Special Case (ii): $M_RN_R>1$, $\delta^{(h)}_{SR}=\delta^{(h)}_{IR}, \delta^{(v)}_{SR}=\delta^{(v)}_{IR}$ and $\eta^{(Q)}>0$;
\item Special Case (iii): $M_RN_R>1$, $\delta^{(h)}_{SR}=\delta^{(h)}_{IR}, \delta^{(v)}_{SR}=\delta^{(v)}_{IR}$ and $\eta^{(Q)} \leq 0$;
\item Special Case (iv): $P_I=0$.
\end{itemize}
\begin{theorem}[Globally Optimal Solutions in Special Cases] \label{lem:op}For $Q=instant$ or $statistic$, the following statements hold. In Special Case (i), any $\boldsymbol{\phi}^{(Q)*}$ satisfying \eqref{eq:phi} is optimal, and $y_{SRU}\left(\boldsymbol\phi^{(Q)*}\right)=y_{IRU}\left(\boldsymbol\phi^{(Q)*}\right)=1$.
In Special Case (ii), any $\boldsymbol\phi^{(Q)*}$ with $\phi^{(Q)*}_{m,n}=\Lambda\left(\alpha-\theta_{IRU,m,n}\right),$ $m \in \mathcal M_R, n \in \mathcal N_R$, for all $\alpha \in \mathbb{R}$, is optimal, and $y_{SRU}\left(\boldsymbol\phi^{(Q)*}\right)=y_{IRU}\left(\boldsymbol\phi^{(Q)*}\right)=M_R^2N_R^2$.
In Special Case (iii), any $\boldsymbol\phi^{(Q)*}$ satisfying $\phi^{(Q)*}_{m,2i}-\phi^{(Q)*}_{m,2i-1}=(2k_i+1)\pi-\left(\theta_{IRU,m,2i}-\theta_{IRU,m,2i-1}\right)$ for some $k_i \in \mathbb{Z}, m \in \mathcal M_R, i=1,...,\frac{N_R}{2}$ and \eqref{eq:phi} is optimal,  and $y_{SRU}\left(\boldsymbol\phi^{(Q)*}\right)=y_{IRU}\left(\boldsymbol\phi^{(Q)*}\right)=0$.
In Special Case (iv), any $\boldsymbol\phi^{(Q)*}$ with $\phi^{(Q)*}_{m,n}=\Lambda\left(\alpha-\theta_{SRU,m,n}\right),$ $m \in \mathcal M_R, n \in \mathcal N_R$, for all $\alpha \in \mathbb{R}$, is optimal, and $y_{SRU}\left(\boldsymbol\phi^{(Q)*}\right)=M_R^2N_R^2$.
\end{theorem}
\begin{IEEEproof}
Please refer to Appendix C.
\end{IEEEproof}
\setcounter{TempEqCnt}{\value{equation}}
\setcounter{equation}{25}
\begin{figure*}
\begin{align}
B^{(Q,t)}_{1,m,n} \triangleq & B^{(Q,t)}_{S,m,n}B^{(Q,t)}_{IRU,m,n}\cos B_{\angle IRU,m,n}^{(t)}-B^{(t)}_{SRU,m,n}B^{(Q,t)}_{I,m,n}\cos B_{\angle SRU,m,n}^{(t)}\label{eq:B1} \\
B^{(Q,t)}_{2,m,n} \triangleq & B^{(Q,t)}_{S,m,n}B^{(Q,t)}_{IRU,m,n}\sin B_{\angle IRU,m,n}^{(t)}-B^{(t)}_{SRU,m,n}B^{(Q,t)}_{I,m,n}\sin B_{\angle SRU,m,n}^{(t)} \label{eq:B2} \\
\overline\phi_{m,n}^{(Q,t)}= &
\begin{cases}
\arctan\frac{B^{(Q,t)}_{1,m,n}}{B^{(Q,t)}_{2,m,n}} -\arccos\frac{B^{(t)}_{SRU,m,n}B^{(Q,t)}_{I,m,n}\sin(B_{\angle SRU,m,n}^{(t)}-B_{\angle IRU,m,n}^{(t)})}{\sqrt{\left(B^{(Q,t)}_{1,m,n}\right)^2+\left(B^{(Q,t)}_{2,m,n}\right)^2}}, &B^{(Q,t)}_{1,m,n}\geq0\\
\arctan\frac{B^{(Q,t)}_{1,m,n}}{B^{(Q,t)}_{2,m,n}} -\arccos\frac{B^{(t)}_{SRU,m,n}B^{(Q,t)}_{I,m,n}\sin(B_{\angle SRU,m,n}^{(t)}-B_{\angle IRU,m,n}^{(t)})}{\sqrt{\left(B^{(Q,t)}_{1,m,n}\right)^2+\left(B^{(Q,t)}_{2,m,n}\right)^2}}+\pi, &B^{(Q,t)}_{1,m,n}<0\label{eq:phimn}
\end{cases}
\end{align}
\hrulefill
\vspace{-1mm}
\end{figure*}
\setcounter{equation}{\value{TempEqCnt}}

Note that based on Theorem~\ref{lem:op}, we can obtain a globally optimal solution in Special Case (iii), by solving a system of linear equations. In addition, substituting $y_{SRU}\left(\boldsymbol\phi^{(Q)*}\right)$ and $y_{IRU}\left(\boldsymbol\phi^{(Q)*}\right)$ into \eqref{eq:ub gamma}, we can obtain the optimal value of Problem~\ref{prob:eq}, i.e., $\gamma_{ub}^{(Q)*}$. Theorem~\ref{lem:op} can be further interpreted as follows.
Statement (i) of Theorem~\ref{lem:op} is for the case of a single-element IRS. In this case, $y_{SRU}(\boldsymbol\phi)=y_{IRU}(\boldsymbol\phi)=1$ for all $\boldsymbol\phi$, and hence the phase shift of the single element has no impact on the average rate or ergodic rate.
 Statement (ii) and Statement (iii) of Theorem 2 are for the symmetric arrangement with $\delta^{(h)}_{SR}=\delta^{(h)}_{IR}$ and $\delta^{(v)}_{SR}=\delta^{(v)}_{IR}$. Accordingly, $y_{SRU}\left(\boldsymbol\phi\right)=y_{IRU}\left(\boldsymbol\phi\right) \triangleq y\left(\boldsymbol\phi\right)$, and $\eta^{(Q)}$ actually represents the derivative of $\gamma^{(Q)}_{ub}(\boldsymbol\phi)$ with respect to $y\left(\boldsymbol\phi\right)$ (please refer to Appendix C for details). When $\eta^{(Q)} > 0$, the phase shifts that achieve the maximum sum channel power of the LoS components of the indirect signal and interference links, i.e., $M_R^2N_R^2$, also maximize the average rate or ergodic rate. When $\eta^{(Q)}<0$, the phase shifts that achieve the minimum sum channel power of the LoS components of the indirect signal and interference links, i.e., $0$, maximize the average rate or ergodic rate.
Statement (iv) of Theorem~\ref{lem:op} is for the case without interference. In this case, the phase shifts that achieve the maximum sum channel power of the LoS components of the indirect links, i.e., $M_R^2N_R^2$, also maximize the average rate or ergodic rate. The optimization result for $Q=instant$ recovers the one under the ULA model for the multi-antenna BS and multi-element IRS in the instantaneous CSI case in \cite{IEEEexample:8746155}.
\subsection{Stationary Point in General Case}\label{lem:generalcase}
In this part, we consider the general case.
Note that the iterative algorithms based on BCD and MM   in \cite{IEEEexample:yu2019enabling,IEEEexample:pan2019intelligent,pan} can be extended to obtain a stationary point of Problem~\ref{prob:eq} in the general case. In particular, in the BCD algorithm, $\phi_{m,n}, m \in \mathcal M_R, n \in \mathcal N_R$ are sequentially updated according to the closed-form optimal solutions of the coordinate optimization problems at each iteration; in the MM algorithm, $\boldsymbol\phi$ are updated according to the closed-form optimal solution of an approximate problem at each iteration. Numerical results show that if $M_RN_R$ is small, the computation time of the BCD algorithm is shorter; otherwise, the computation time of the MM algorithm is shorter. As neither the BCD algorithm nor the MM algorithm allows parallel computation, their computation efficiencies
on a multi-core processor may be low, especially when $M_RN_R$ is large.
In the following, we propose an iterative algorithm based on PCD, where at each iteration, $\phi_{m,n}, m \in \mathcal M_R, n \in \mathcal N_R$ are updated in parallel, each according to a closed-form expression, to obtain a stationary point of Problem~\ref{prob:eq}. The goal is to improve computation efficiency when multi-core processors are available, especially for large $M_RN_R$.
Let $\boldsymbol\phi^{(t)} \triangleq \left(\phi_{m,n}^{(t)}\right)_{m \in \mathcal M_R, n \in \mathcal N_R}$ denote the phase shifts at the $t$-th iteration. At each iteration, we first maximize $\gamma_{ub}^{(Q)}(\boldsymbol\phi)$ w.r.t. each phase shift $\phi_{m,n}$ with the other phase shifts being fixed.

\begin{Prob}[Block-wise Optimization Problem w.r.t. $\phi_{m,n}$ at Iteration $t$]\label{prob:app}
\begin{align*}
\overline\phi_{m,n}^{\left(Q,t\right)}\!\triangleq\!
\mathop{\arg\!\max}_{\boldsymbol\phi} &\frac{B^{(t)}_{SRU,m,n}\!\cos(\phi_{m,n}\!+\!B_{\angle SRU,m,n}^{(t)})\!+\!B^{(Q,t)}_{S,m,n}}{B^{(Q,t)}_{IRU,m,n}\!\cos(\phi_{m,n}\!+\!B_{\angle IRU,m,n}^{(t)})\!+\!B^{(Q,t)}_{I,m,n}}\!, \\
s.t. &\quad \eqref{eq:phi},
\end{align*}
where
\begin{align*}
B^{(t)}_{SRU,m,n}\triangleq & 2A_{SRU,LoS}\left\lvert\sum\limits_{k \neq m,l\neq n}e^{j\left(\phi_{k,l}^{(t)}+\theta_{SRU,k,l}\right)}\right\rvert,\\
B^{(Q,t)}_{S,m,n}\triangleq &
A_{SRU,LoS}\left(1+
\left\lvert\sum\limits_{k \neq m,l\neq n}e^{j\left(\phi_{k,l}^{(t)}+\theta_{SRU,k,l}\right)}\right\rvert^2\right)
\\ &+A_{SRU,NLoS}^{(Q)}+A_{SU}^{(Q)},\\
B^{(Q,t)}_{IRU,m,n}\triangleq &
2A_{IRU,LoS}^{(Q)}\left\lvert\sum\limits_{k \neq m,l\neq n}e^{j\left(\phi_{k,l}^{(t)}+\theta_{IRU,k,l}\right)}\right\rvert,
\\
B^{(Q,t)}_{I,m,n}\triangleq&
A_{IRU,LoS}^{(Q)}\left(1+
\left\lvert\sum\limits_{k \neq m,l\neq
n}e^{j\left(\phi_{k,l}^{(t)}+\theta_{IRU,k,l}\right)}\right\rvert^2\right)
\\ &+A_{IRU,NLoS}^{(Q)}+A_{IU},
\end{align*}
\begin{align*}
B_{\angle cRU,m,n}^{(t)}\triangleq &\theta_{cRU,k,l}-\angle\left(\sum\limits_{k \neq m,l\neq n}e^{j\left(\phi_{k,l}^{(t)}+\theta_{cRU,k,l}\right)}\right)
.
\end{align*}
\end{Prob}

By taking the derivative of the objective function of Problem~\ref{prob:app} w.r.t. $\phi_{m,n}$, and setting it to zero, we obtain the following equation:
\begin{align}
&B^{(Q,t)}_{1,m,n}\sin(\phi_{k,l})+B^{(Q,t)}_{2,m,n}\cos(\phi_{k,l})
\nonumber \\ =& B^{(t)}_{SRU,m,n}B^{(Q,t)}_{IRU,m,n} \sin(B_{\angle SRU,m,n}^{(t)}-B_{\angle IRU,m,n}^{(t)}),\label{eq:deri}
\end{align}where $B^{(Q,t)}_{1,m,n}$ and $B^{(Q,t)}_{2,m,n}$ are given by \eqref{eq:B1} and \eqref{eq:B2}, as shown at the top of the page.
The equation in \eqref{eq:deri} has two possible roots. By further checking the second derivative of the objective function of Problem 2, we obtain the closed-form optimal solution of Problem~\ref{prob:app} in \eqref{eq:phimn}, as shown at the top of the page. Then, we update $\boldsymbol\phi^{(t+1)}$ according to:
\setcounter{TempEqCnt}{\value{equation}}
\setcounter{equation}{29}
\begin{figure*}
 \begin{align}
 \zeta^{(Q)}\left(\boldsymbol\phi^*\right) \leq & \log_2\left(\frac{1+\frac{A_{SRU,LoS}+A_{SRU,NLoS}^{(Q)}
+A_{SU}^{(Q)}}
{A_{IRU,LoS}^{(Q)}+A_{IRU,NLoS}^{(Q)}+A_{IU}}}{1+\frac{4\lceil \frac{M_RN_R-1}{2} \rceil^2A_{SRU,LoS}
\cos^2\frac{2\pi}{2^{b+1}}+A_{SRU,NLoS}^{(Q)}
+A_{SU}^{(Q)}}{4\lceil \frac{M_RN_R-1}{2} \rceil^2A_{IRU,LoS}^{(Q)}
\cos^2\frac{2\pi}{2^{b+1}}+A_{IRU,NLoS}^{(Q)}+A_{IU}}}\right) \label{eq:q1} \\
\zeta^{(Q)}\left(\boldsymbol\phi^*\right)\leq & \log_2\left(\frac{1+\frac{A_{SRU,NLoS}^{(Q)}
+A_{SU}^{(Q)}}
{A_{IRU,NLoS}^{(Q)}+A_{IU}}}{1+\frac{4\lceil \frac{M_RN_R-1}{2} \rceil^2A_{SRU,LoS}
\sin^2\frac{2\pi}{2^{b+1}}+A_{SRU,NLoS}^{(Q)}
+A_{SU}^{(Q)}}{4\lceil \frac{M_RN_R-1}{2} \rceil^2A_{IRU,LoS}^{(Q)}
\sin^2\frac{2\pi}{2^{b+1}}+A_{IRU,NLoS}^{(Q)}+A_{IU}}}\right)
\label{eq:q2}\\
\zeta^{(Q)}\left(\boldsymbol\phi^*\right)\leq &
\log_2\left(\frac{\sigma^2+A_{SRU,LoS}+A_{SRU,NLoS}^{(Q)}
+A_{SU}^{(Q)}}
{\sigma^2+4\lceil \frac{M_RN_R-1}{2} \rceil^2A_{SRU,LoS}
\cos^2\frac{2\pi}{2^{b+1}}+A_{SRU,NLoS}^{(Q)}
+A_{SU}^{(Q)}}\right) \label{eq:q4} \\
\zeta^{(Q)}\left(\boldsymbol\phi^\dag\right) \leq &\frac{2\pi M_RN_R\left\lvert A_{IRU,LoS}^{(Q)}\left(A_{SRU,NLoS}^{(Q)}+
A_{SU}^{(Q)}\right)-A_{SRU,LoS}\left(
A_{IRU,NLoS}^{(Q)}+A_{IU}\right)\right\rvert(M_RN_R-1)}
{2^b\ln2\left(A_{IRU,NLoS}^{(Q)}+A_{IU}\right)
\left(A_{SRU,NLoS}^{(Q)}+A_{SU}^{(Q)}+A_{IRU,NLoS}^{(Q)}+A_{IU}\right)}\label{eq:q3}
\end{align}
\hrulefill
\vspace{-1mm}
\end{figure*}
\setcounter{equation}{\value{TempEqCnt}}
\setcounter{equation}{28}
\begin{align}
&\phi_{m,n}^{(Q,t+1)}=(1-\rho^{(t)})\phi_{m,n}^{(Q,t)}
+\rho^{(t)}\overline\phi_{m,n}^{(Q,t)},\label{eqn:updatephi}
\end{align}
where $m \in \mathcal M_R, n \in \mathcal N_R$ and $\rho^{(t)}$ is a positive diminishing stepsize satisfying:
\begin{align*}
&\rho^{(t)}>0,\ \lim_{t\to\infty}\rho^{(t)}=0,\ \sum_{t=1}^\infty\rho^{(t)}=\infty,\ \sum_{t=1}^\infty\left(\rho^{(t)}\right)^2<\infty. \label{eqn:gamma}
\end{align*}
The details of the PCD algorithm are summarized in Algorithm~\ref{alg:BCD}.\footnote{Algorithm~\ref{alg:BCD} is suitable for the cases which are not covered in Theorem~\ref{lem:ergodic Case 1 with reflector}.} By \cite{IEEEexample:PCD}, we know that $\boldsymbol\phi^{(t)}\to\boldsymbol\phi^{+}$ as $t\to\infty$, where $\boldsymbol\phi^{+}$ is a stationary point of Problem~\ref{prob:eq}.
\begin{algorithm}[t]\small
    \caption{PCD Algorithm for Obtaining a Stationary Point in General Case}
\begin{small}
        \begin{algorithmic}[1]
           \STATE \textbf{initialization}: choose any $\boldsymbol\phi^{(Q,0)}$ as the initial point, and set $t=0$.\\
           \STATE \textbf{repeat}
           \STATE \quad For all $m \in \mathcal M_R$ and $n \in \mathcal N_R$, compute $\overline\phi_{m,n}^{(Q,t)}$ according to \eqref{eq:phimn}.
           \STATE \quad Update $\boldsymbol\phi^{(Q,t+1)}$ according to \eqref{eqn:updatephi}.
           \STATE\quad Set $t=t+1$.
           \STATE \textbf{until} some convergence criterion is met.
    \end{algorithmic}\label{alg:BCD}
    \end{small}
\end{algorithm}
\subsection{Quantization}\label{Quantization}
In practice, the phase shift design is subject to quantization error. We consider a uniform scalar quantizer with $b$ quantization bits \cite{IEEEexample:8811733,IEEEexample:8746155}. Then, for all $m \in \mathcal M_R$ and $n \in \mathcal N_R$, the quantization error for the phase shift of the $(m,n)$-th element, denoted by $\delta_{m,n}$, lies in $\left[-\frac{2\pi}{2^{b+1}},\frac{2\pi}{2^{b+1}}\right]$. Denote $\boldsymbol\delta \triangleq \left(\delta_{m,n}\right)_{m \in \mathcal M_R, n \in \mathcal N_R}$. Let $\zeta^{(Q)}\boldsymbol(\boldsymbol\phi)\triangleq C_{ub}^{(Q)}(\boldsymbol\phi)-C_{ub}^{(Q)}
(\boldsymbol\phi+\boldsymbol\delta)$ denote the average or ergodic rate degradation at the phase shifts $\boldsymbol\phi$ due to quantization. The following theorem shows the average rate degradation and the ergodic rate degradation at the optimal solutions in the four special cases and a stationary point in the general case.
\begin{theorem}\label{lem:quantize}
(i): In Special Case (i), $\zeta^{(Q)}(\boldsymbol\phi)=0$.
In Special Case (ii), Special Case (iii) and Special Case (iv), the upper bounds of $\zeta^{(Q)}\left(\boldsymbol\phi^*\right)$ are given by \eqref{eq:q1}, \eqref{eq:q2} and \eqref{eq:q4}, respectively, as shown at the top of this page.
(ii): In the general case, the upper bound of $\zeta^{(Q)}\left(\boldsymbol\phi^*\right)$ is given by \eqref{eq:q3}, as shown at the top of the page.
(iii): The upper bounds in \eqref{eq:q1}, \eqref{eq:q2}, \eqref{eq:q4} and \eqref{eq:q3} decrease with $b$.
\end{theorem}
\begin{IEEEproof}
Please refer to Appendix D.
\end{IEEEproof}

As $b \rightarrow \infty$, the upper bounds in Theorem~\ref{lem:quantize} go to zero. That is, the upper bounds are asymptotically  tight at large $b$.
\section{Comparision with System without IRS}\label{section:comparision}
In this section, to characterize the benefit of IRS in downlink transmission with interference, we first present a counterpart system without IRS, and analyze its average rate in the instantaneous CSI case and ergodic rate in the statistical CSI case. Then, we compare them with those of the IRS-assisted system.
\subsection{System without IRS}
In the counterpart system without IRS, the signal received at user $U$ is expressed as:
\setcounter{equation}{33}
\begin{align}
Y_{no} \triangleq \sqrt{P_S}\mathbf{h}^H_{SU} \mathbf{w}_{no,S} X_S + \sqrt{P_I} \mathbf{h}^H_{IU} \mathbf{w}_{no,I} X_I +Z,
\end{align}
where $\mathbf{w}_{no,S}$ and $\mathbf{w}_{no,I}$ denote the beamforming vectors for the signal BS and interference BS, respectively, satisfying $\lvert\lvert\mathbf{w}_{no,S}\rvert\rvert^2_2=1$ and $\lvert\lvert\mathbf{w}_{no,I}\rvert\rvert^2_2=1$. Analogously, assume that user $U$ knows $\mathbf{h}^H_{SU} \mathbf{w}_{no,S}$, but does not know $\mathbf{h}^H_{IU} \mathbf{w}_{no,I}$. In the following, we consider the instantaneous CSI case and the statistical CSI case, respectively.
\setcounter{TempEqCnt}{\value{equation}}
\setcounter{equation}{38}
\begin{figure*}[t]
\begin{align}
\varsigma^{(instant)}\triangleq&
\alpha_{IU}\left(
A_{SRU,LoS}M_R^2N_R^2+A_{SRU,NLoS}^{(instant)}\right)-A_{SU}^{(instant)}
\alpha_{IR}\alpha_{RU}\left(\tau_{IRU}M_R^2N_R^2  +M_RN_R(1-\tau_{IRU})
\right)\label{eq:coPI} \\
\varsigma^{(statistic)}\triangleq& \alpha_{IU}\left(
A_{SRU,LoS}M_R^2N_R^2+A_{SRU,NLoS}^{(statistic)}\right)
-A_{SU}^{(statisic)}
\alpha_{IR}\alpha_{RU}\left(\frac{\tau_{IRU}y_{IR}}{M_IN_I}
\!+ \! \frac{M_RN_R\left(M_IN_IK_{RU}+\tau_Iy_{IR}\right)}
{M_IN_IK_{RU}(K_{IR}+1)}\right)
\label{eq:coPI1}
\end{align}
\hrulefill
\vspace{-1mm}
\end{figure*}
\setcounter{equation}{\value{TempEqCnt}}
\subsubsection{Instantaneous CSI Case}
In this part, assume that the CSI of the channel between the signal BS and user $U$, i.e., $\mathbf{h}^H_{SU}$, is known at the signal BS and the CSI of the channel between the interference BS and user $U'$, i.e., $\mathbf{h}^H_{IU}$, is known at the interference BS. Consider the instantaneous CSI-adaptive MRT at the signal BS and interference BS, respectively, i.e.,
$\mathbf{w}^{(instant)}_{no,S} =\frac{\mathbf{h}_{SU}}{\left\lvert\left\lvert \mathbf{h}_{SU}\right\rvert\right\rvert_2}$ and $\mathbf{w}^{(instant)}_{no,I} = \frac{\mathbf{h}_{IU'}}{\left\lvert\left\lvert \mathbf{h}_{IU'}\right\rvert\right\rvert_2}$.
Then, the average rate of the counterpart system without IRS is given by:
\begin{align}
C_{no}^{(instant)}\! = \! \mathbb{E}\!\left[\!\log_2\!\left(\!1+ \! \frac{P_S\alpha_{SU} {\left\lvert\left\lvert\mathbf{h}_{SU}\right\rvert\right\rvert}_2^2}{P_I\alpha_{IU} \mathbb{E}\left[\left\lvert \mathbf{h}_{IU}^H\frac{\mathbf{h}_{IU'}}{\left\lvert\left\lvert
\mathbf{h}_{IU'}\right\rvert\right\rvert_2}\right\rvert^2\right]+ \sigma^2}\!\right)\!\right]\!.
\end{align}
Similarly, for tractability, we can obtain an analytical upper bound of $C_{no}^{(instant)}$, i.e.,
$C_{no}^{(instant)} \leq \log_2(1+\gamma^{(instant)}_{no,ub}) \triangleq C^{(instant)}_{no,ub}$,
where $
\gamma^{(instant)}_{no,ub} \triangleq \frac{A_{SU}^{(instant)}}{A_{IU}}$.
\subsubsection{Statistical CSI Case}
In this part, assume that the BSs have no channel knowledge. We consider isotropic transmission at the signal BS and interference BS, i.e., $\mathbf{w}_{no,S}^{(statistic)} = \frac{1}{\sqrt{M_SN_S}} \boldsymbol{1}_{M_SN_S} \in \mathbb{C}^{M_SN_S \times 1}$ and
$\mathbf{w}_{no,I}^{(statistic)} = \frac{1}{\sqrt{M_IN_I}} \boldsymbol{1}_{M_IN_I} \in \mathbb{C}^{M_IN_I \times 1}$.
Then,  coding over a large
number of channel coherence time intervals, the ergodic rate of the counterpart system without IRS is given by:
\begin{align}
C_{no}^{(statistic)} \!=\!  \mathbb{E}\!\left[\!\log_2\!\left(\!1+ \! \frac{\frac{P_S\alpha_{SU}}{M_SN_S} { \left\lvert\mathbf{h}_{SU}^H\boldsymbol{1}_{M_SN_S}\right\rvert }^2}{\frac{P_I\alpha_{IU}}{M_IN_I}\mathbb{E}\left[{\left\lvert\mathbf{h}^H_{IU} \boldsymbol{1}_{M_IN_I}\right\rvert}^2\right]+ \sigma^2}\!\right)\!\right]\!.
\end{align}
Similarly, we can obtain an analytical upper bound of $C_{no}^{(statistic)}$, i.e.,
$C_{no}^{(statistic)} \leq \log_2\left(1\right.$ $\left.+\gamma^{(statistic)}_{no,ub}\right)\triangleq C^{(statistic)}_{no,ub}$,
where $
\gamma^{(statistic)}_{no,ub} \triangleq \frac{A_{SU}^{(statistic)}}{A_{IU}}$.
\subsection{Comparision}
In this part, we compare $\gamma_{ub}^{(Q)}\left(\phi^*\right)$ and $\gamma_{no,ub}^{(Q)}$. For $Q=instant$ or $statistic$, define:
\begin{align}
\xi^{(Q)}_{>}\triangleq &
\left(A_{SRU,LoS}A_{IU}-
A_{IRU,LoS}^{(Q)}A_{SU}^{(Q)}\right)M_R^2N_R^2
\nonumber\\&+A_{SRU,NLoS}^{(Q)}
A_{IU}-A_{SU}^{(Q)}A_{IRU,NLoS}^{(Q)} ,
\label{eq:xi1}
\end{align}
\begin{align}
\xi^{(Q)}_{<}\triangleq&
A_{SRU,LoS}A_{IU}M_R^2N_R^2
\nonumber \\ &+A_{SRU,NLoS}^{(Q)}
A_{IU}-A_{SU}^{(Q)}A_{IRU,NLoS}^{(Q)}.
\label{eq:xi2}
\end{align}By comparing \eqref{eq:xi1} and \eqref{eq:xi2}, it is clear that $\xi^{(Q)}_{>} < \xi^{(Q)}_{<}$, for $Q=instant$ or $statistic$.
\begin{theorem}[Comparision]\label{lem:extreme cases1}For $Q=instant$ or $statistic$, the following statements hold. If $\xi^{(Q)}_{>} > 0$,
then $\gamma^{(Q)}_{ub}(\boldsymbol\phi^{*})>\gamma^{(Q)}_{no,ub}$ ;
if $\xi^{(Q)}_{<} <0$,
then $\gamma^{(Q)}_{ub}(\boldsymbol\phi^{*})<\gamma^{(Q)}_{no,ub}$.
\end{theorem}
\begin{IEEEproof}
Please refer to Appendix E.
\end{IEEEproof}

From \eqref{eq:xi1} and \eqref{eq:xi2}, we know that $\xi^{(Q)}_{>}$ and $\xi^{(Q)}_{<}$ increase with $\alpha_{SR}$, $\alpha_{IU}$ and $\tau_{SRU}$ and decrease with $\alpha_{IR}$, $\alpha_{SU}$ and $\tau_{IRU}$.
Thus, from Theorem~\ref{lem:extreme cases1}, we can draw the following conclusions. If the channel between the signal BS and the IRS is strong, the interference BS and the IRS is weak, the channel between the interference BS and user $U$ is strong, the signal BS and user $U$ is weak, the LoS components of the indirect link between the signal BS and user $U$ via the IRS are dominant, or the interference BS and user $U$ via the IRS are not dominant, the IRS-assisted system with the optimal quasi-static phase shift design is effective for improving the average rate in the instantaneous CSI case and the ergodic rate in the statistical CSI case, in the presence of interference. Otherwise, the system without IRS is beneficial in the presence of interference.
Define $\varsigma^{(instant)}$ and $\varsigma^{(statistic)}$ in \eqref{eq:coPI} and \eqref{eq:coPI1}, as shown at the top of the page.
From Theorem~\ref{lem:extreme cases1}, we have the following corollary.
\begin{corollary}\label{cor:compare}
For $Q=instant$ or $statistic$, the following statements hold. If $\varsigma^{(Q)} > 0$, then $\gamma^{(Q)}_{ub}(\boldsymbol\phi^{*})>\gamma^{(Q)}_{no,ub}$; if $\varsigma^{(Q)} < 0$ and $P_I \leq \varepsilon$ for some $\varepsilon > 0$, then $\gamma^{(Q)}_{ub}(\boldsymbol\phi^{*})>\gamma^{(Q)}_{no,ub}$;
if   $\varsigma^{(Q)} < 0$ and $P_I> \varepsilon$ for some $\varepsilon >0$, then
$\gamma^{(Q)}_{ub}(\boldsymbol\phi^{*})<\gamma^{(Q)}_{no,ub}$.
\end{corollary}
\begin{IEEEproof}
Substituting \eqref{eq:A4}, \eqref{eq:A5} and \eqref{eq:A6} into \eqref{eq:xi1}, we have $\xi_{>}^{(Q)}=P_I\varsigma^{(Q)}+\sigma^2\left(
A_{SRU,LoS}\right. $ $\left.M_R^2N_R^2+A_{SRU,NLoS}^{(Q)}\right)$. Note that $\sigma^2$ $\left(A_{SRU,LoS}M_R^2N_R^2+A_{SRU,NLoS}^{(Q)}\right)>0$.
Thus, if $\varsigma^{(Q)} > 0$, then $\gamma^{(Q)}_{ub}(\boldsymbol\phi^{*})>\gamma^{(Q)}_{no,ub}$; if $\varsigma^{(Q)} < 0$, $\varsigma^{(Q)}_{>}$ decreases with $P_I$. Therefore, by Theorem~\ref{lem:extreme cases1}, we can complete the proof of Corollary~\ref{cor:compare}.
\end{IEEEproof}

From \eqref{eq:coPI} and \eqref{eq:coPI1}, we know that $\varsigma^{(Q)}$ increases with $\alpha_{SR}$, $\alpha_{IU}$ and $\tau_{SRU}$ and decrease with $\alpha_{IR}$, $\alpha_{SU}$ and $\tau_{IRU}$. Thus, from Corollary~\ref{cor:compare}, we can make the following conclusions. If the channel between the signal BS and the IRS is strong, or the interference BS and the IRS is weak, the channel between the interference BS and user $U$ is strong, or the signal BS and user $U$ is weak, the LoS components of the indirect link between the signal BS and user $U$ via the IRS are dominant, or the interference BS and user $U$ via the IRS are not dominant, the IRS-assisted system with the optimal quasi-static phase shift design is effective at any $P_I$. Otherwise, it is effective only if $P_I$ is small enough.  Furthermore, if $P_I=0$, the IRS-assisted system with the optimal quasi-static phase shift design is always beneficial.

\section{Numerical Results}\label{sec:simulation}
\begin{figure}[t]
\begin{center}
\includegraphics[width=6cm]{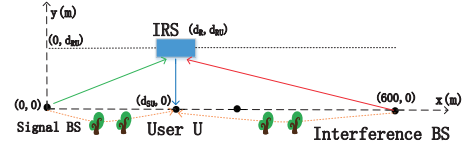}
\end{center}
\vspace{-3mm}
\caption{\small{The IRS-assisted system considered in Section \ref{sec:simulation} \cite{IEEEexample:pan2019intelligent}.}}
\label{fig:pathloss}
\vspace{-3mm}
\end{figure}

In this section, we numerically evaluate the performance of the proposed solutions in an IRS-assisted system~\cite{IEEEexample:pan2019intelligent}, where the signal BS, the interference BS, user $U$ and the IRS are located at $(0,0)$, $(600,0)$, $(d_{SU},0)$, $(d_{R},d_{RU})$ (in m), respectively, and user $U$ lies on the line between the signal BS and the interference BS, as shown in Fig. \ref{fig:pathloss}.
In the simulation, we set $d=\frac{\lambda}{2}$, $M_S=N_S=4$, $M_I=N_I=4$, $M_R=N_R=8$, $P_S=P_I=30$dBm, $\sigma^2=-104$dBm, $\varphi_{SR}^{(h)}=\varphi_{SR}^{(v)}=\pi/3$, $\varphi_{IR}^{(h)}=\varphi_{IR}^{(v)}=\pi/8$, $\varphi_{RU}^{(h)}=\varphi_{RU}^{(v)}=\pi/6$, $d_R=250$m, $d_{SU}=250$m, $d_{RU}=20$m, if not specified otherwise.
We consider the path loss model in \cite{IEEEexample:guo2019weighted,
IEEEexample:8811733,IEEEexample:pan2019intelligent}, and choose similar path loss exponents to those in \cite{IEEEexample:guo2019weighted,
IEEEexample:8811733,IEEEexample:pan2019intelligent}.
Specifically, the distance-dependent path losses $\alpha_{SU}$, $\alpha_{IU}$, $\alpha_{SR}$, $\alpha_{IR}$, $\alpha_{RU}$ follow
$\alpha_i=\frac{1}{1000 d_i^{\bar\alpha_i}}$ $(\text{i.e.,}-30+10\bar\alpha_i\log_{10}(d_i) \text{ dB}),\ i=SU, IU, SR, IR, RU$ \cite{IEEEexample:guo2019weighted,
IEEEexample:8811733,IEEEexample:pan2019intelligent}.
Due to extensive obstacles and scatters, we set $\bar\alpha_{SU}=3.7$ and $\bar\alpha_{IU}=3.5$.
As the location of the IRS is usually carefully chosen, we assume that the links between the BSs and the IRS experience free-space path loss, and set $\bar\alpha_{SR}=\bar\alpha_{IR}=2$, as in \cite{IEEEexample:guo2019weighted}. In addition, we set $\bar\alpha_{RU}=3$, due to few obstacles.
\begin{figure}[t]
\begin{center}
\subfigure[\scriptsize{Instantaneous CSI case.
}\label{fig:1a}]
{\resizebox{4.25cm}{!}{\includegraphics{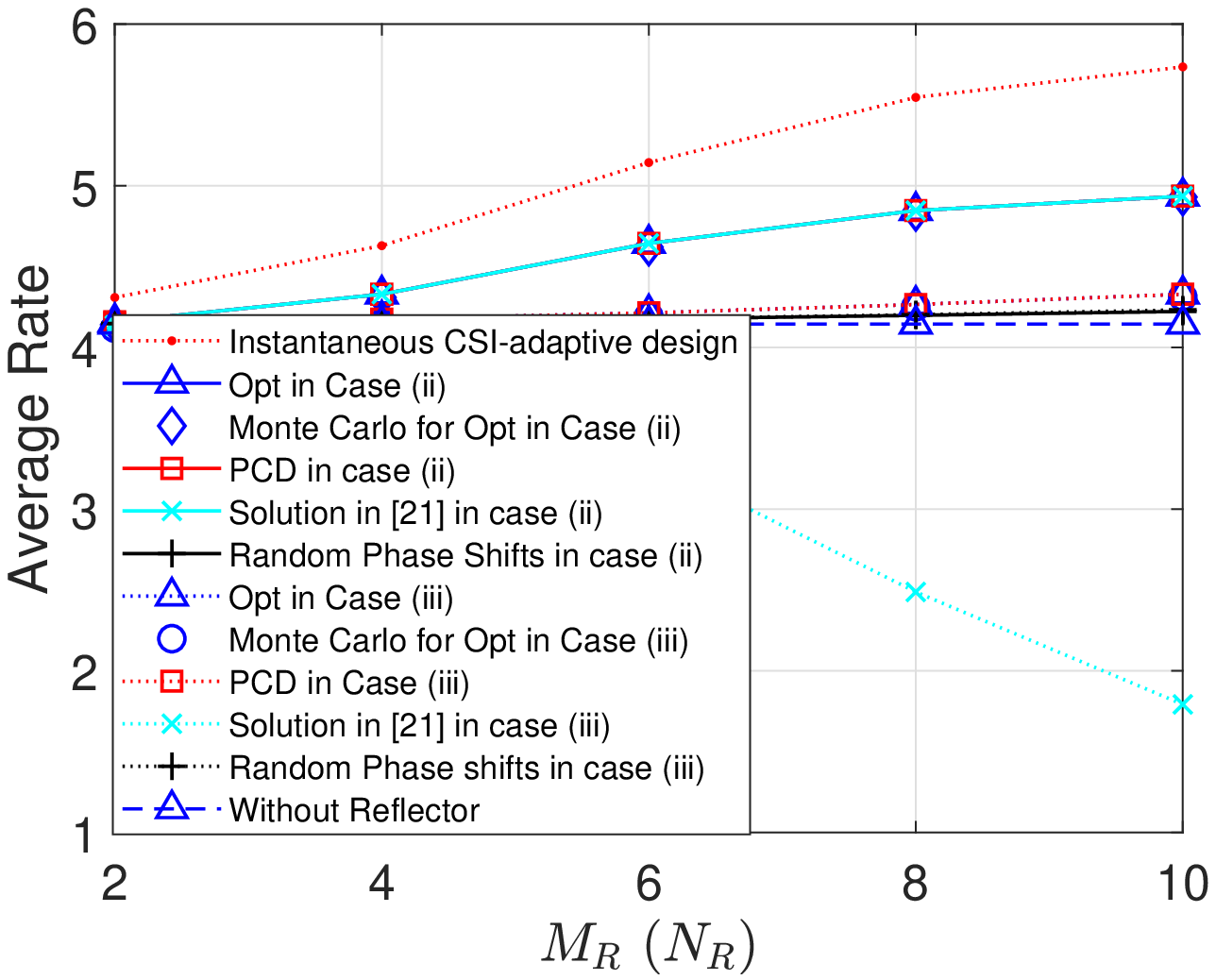}}}\quad
\subfigure[\scriptsize{Statistical CSI case.
}\label{fig:1b}]
{\resizebox{4.25cm}{!}{\includegraphics{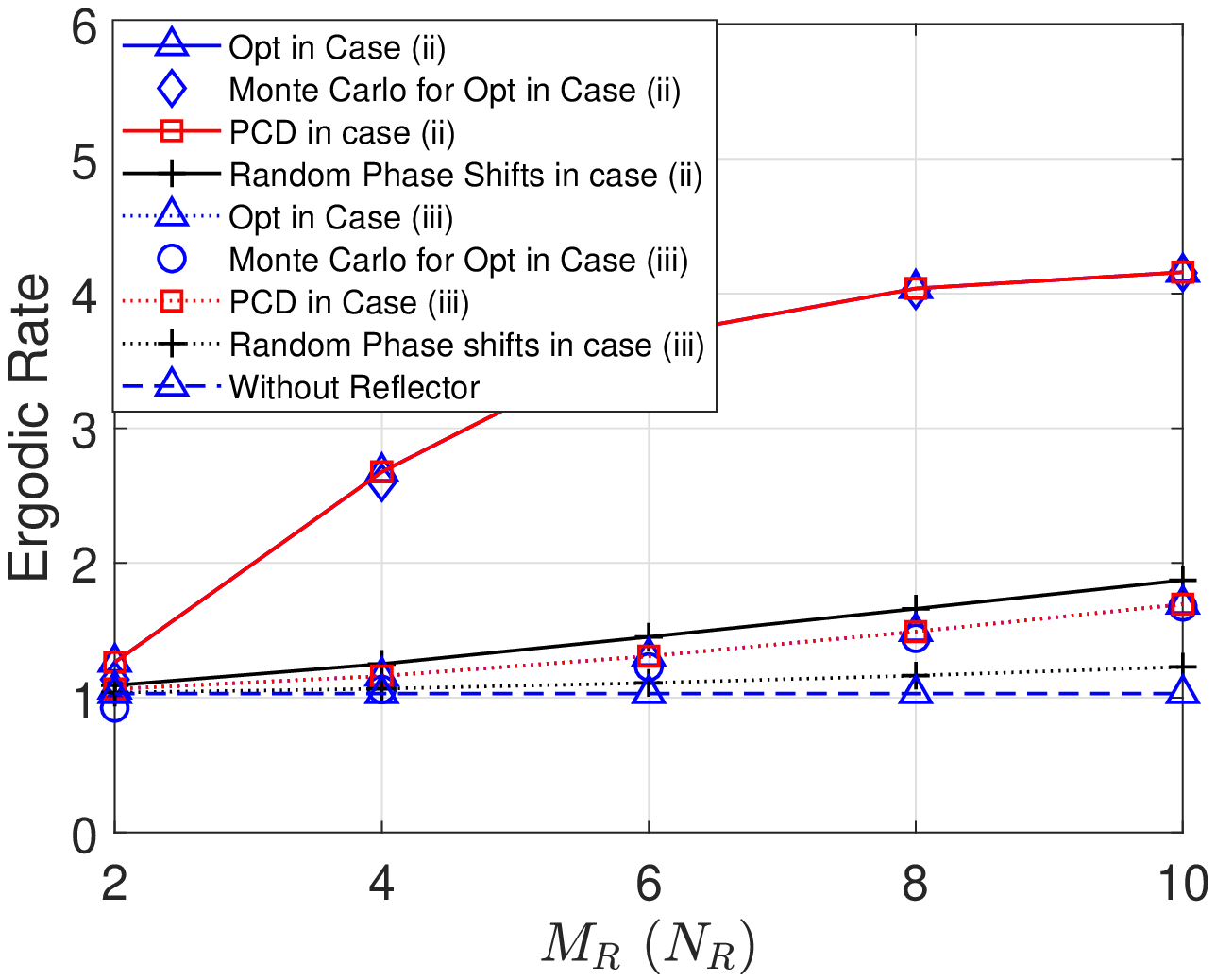}}}
\end{center}
\vspace{-2mm}
\caption{\small{Average rate and ergodic rate versus $M_R$ $(=\!N_R)$ in special cases.}}
\vspace{-2mm}
\label{fig:optimal_M_R}
\end{figure}
\begin{figure}[t]
\begin{center}
\subfigure[\scriptsize{Instantaneous CSI case.
}\label{fig:2a}]
{\resizebox{4.25cm}{!}{\includegraphics{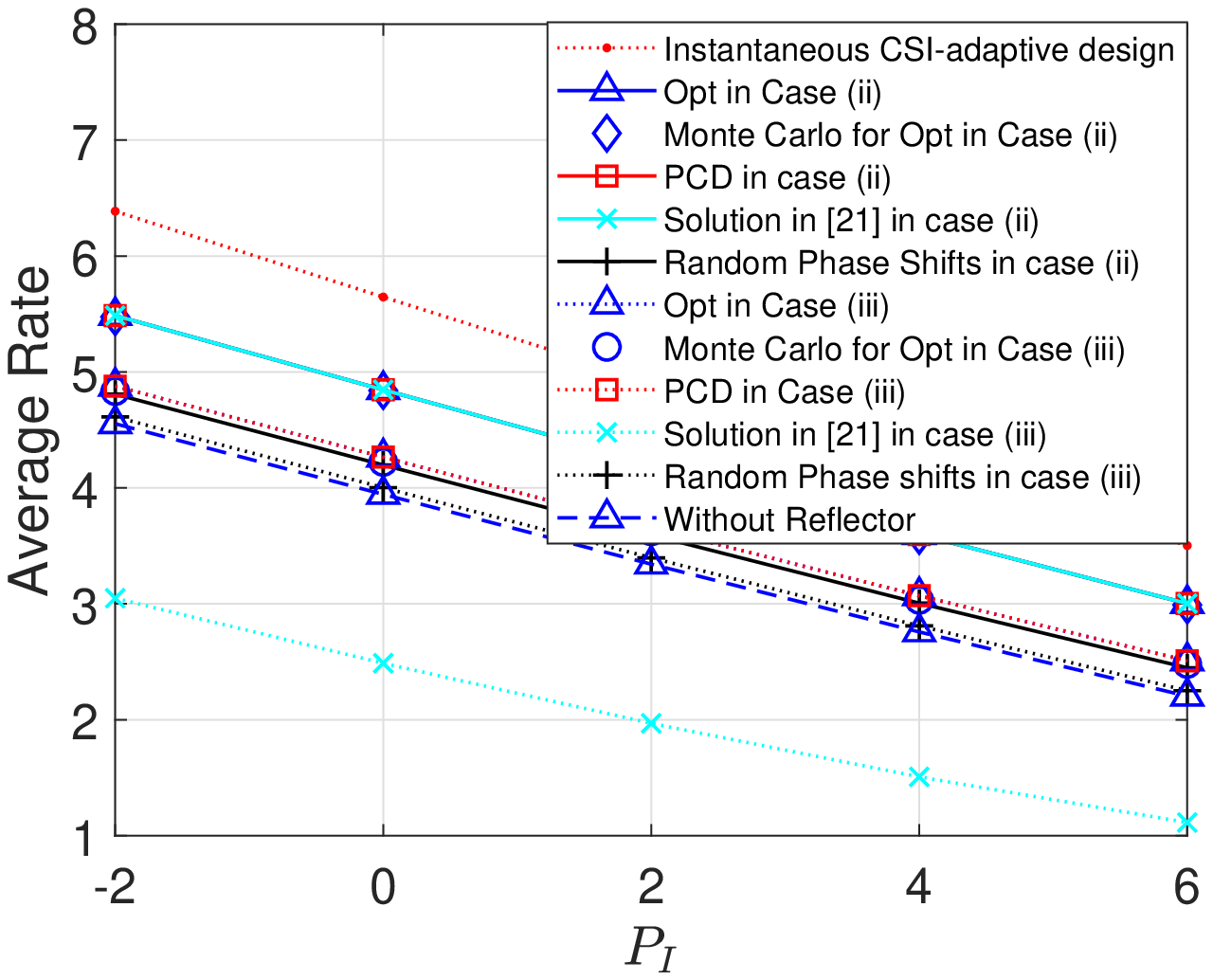}}}\quad
\subfigure[\scriptsize{Statistical CSI case.
}\label{fig:2b}]
{\resizebox{4.25cm}{!}{\includegraphics{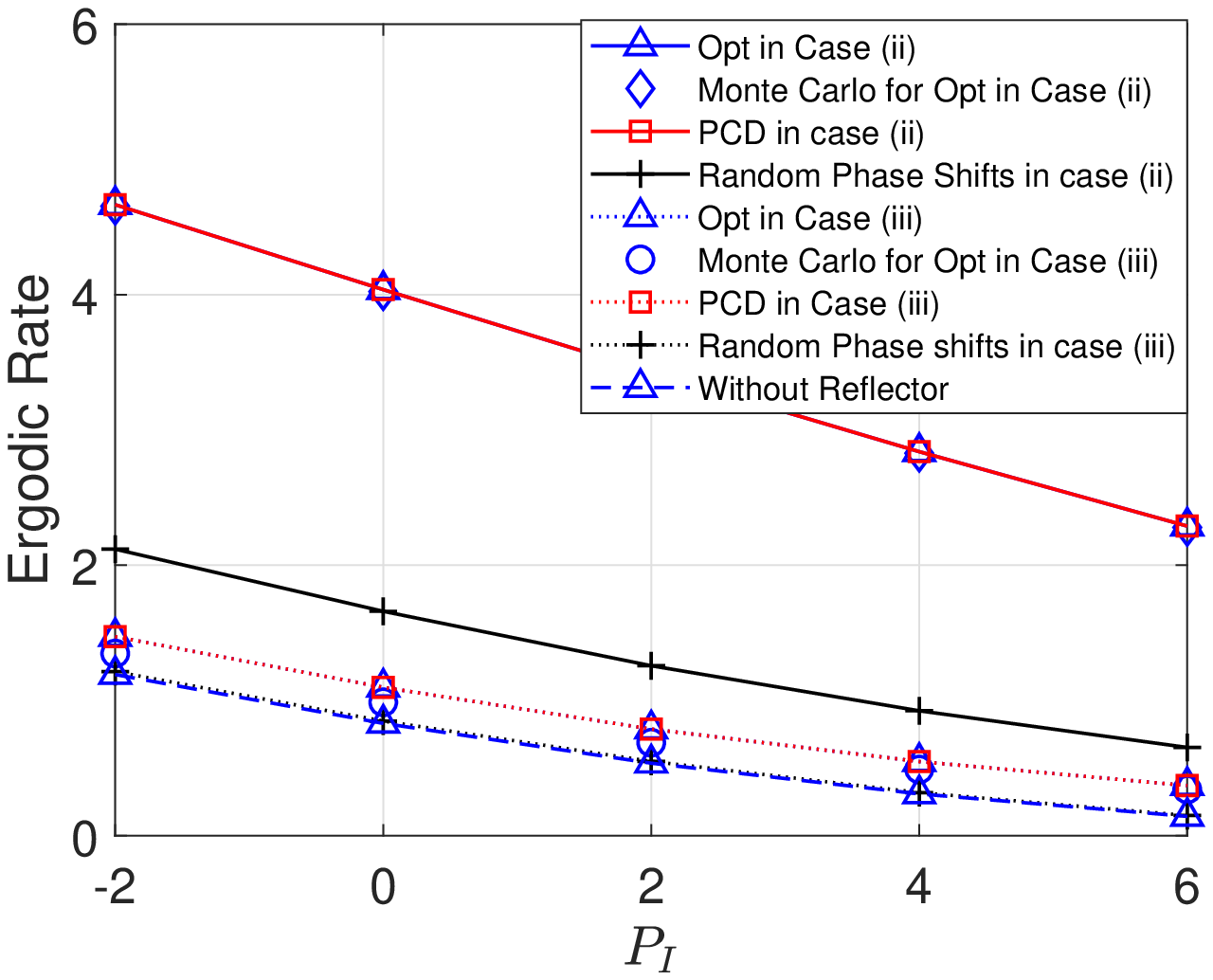}}}
\end{center}
\vspace{-2mm}
\caption{\small{Average rate and ergodic rate versus $P_I$ in special cases.}}
\vspace{-2mm}
\label{fig:optimal_P_I}
\end{figure}

We consider four baseline schemes. Baseline 1 and Baseline~2 are applicable for both the instantaneous CSI case and the statistical CSI case. In contrast, Baseline 3 and Baseline 4 are applicable only for the instantaneous CSI case. In particular, Baseline 1 reflects the average rate and ergodic rate of the counterpart system without IRS in Section~\ref{section:comparision} \cite{IEEEexample:8811733,IEEEexample:guo2019weighted,
IEEEexample:zhang2019analysis}; Baseline 2 chooses the phase shifts uniformly at random \cite{IEEEexample:8811733,IEEEexample:guo2019weighted,IEEEexample:8746155}, and shows the average rate and ergodic rate obtained by averaging over $10000$ random choices;
Baseline 3 implements the phase shifts $\boldsymbol\phi_{opt} \triangleq \left(\phi_{opt,m,n}\right)_{m \in \mathcal M_R, n \in \mathcal N_R}$ with $\phi_{opt,m,n}=\Lambda\left(\alpha-\theta_{SRU,m,n}\right)$, which maximize the received signal power (without considering interference); Baseline 4 is the instantaneous CSI-adaptive phase shift design corresponding to a stationary point of the maximization problem of $\gamma^{(instant)}(\boldsymbol\phi)$ in \eqref{eq:sinr} subject to the constraints in \eqref{eq:phi}, which is obtained by a PCD algorithm similar to Algorithm~\ref{alg:BCD}. Note that Baseline 3 is an extension of the optimal solution for the instantaneous CSI case under the ULA model in \cite{IEEEexample:8746155} to the URA model. In addition, it is worth noting that Baseline 4 achieves the maximum average rate in the instantaneous CSI case, with the highest phase adjustment cost. In the general case, besides the proposed PCD algorithm, we also evaluate the BCD and MM algorithms\cite{IEEEexample:yu2019enabling}.  We adopt the same convergence criterion, i.e., $\gamma_{ub}^{(Q)}\left(\boldsymbol\phi^{(t+1)}_{opt}\right)-
\gamma_{ub}^{(Q)}\left(\boldsymbol\phi^{(t)}_{opt}\right) \leq 10^{-6}$, for the PCD, BCD and MM algorithms. For ease of illustration, we refer to the stationary points obtained by the PCD, BCD and MM algorithms as the PCD, BCD and MM solutions, respectively.

We set $\delta_{SR}^{(h)}=\delta_{SR}^{(v)}=\pi/6$, $\delta_{IR}^{(h)}=\delta_{IR}^{(v)}=\pi/6$ in Special Case (ii) and Special Case (iii), set $K_{SR}=K_{IR}=K_{RU}=20$dB in Special Case (ii), and set $K_{SR}=-20$dB, $K_{IR}=K_{RU}=20$dB in Special Case (iii).
Fig. \ref{fig:optimal_M_R} and Fig. \ref{fig:optimal_P_I} illustrate the average rate and ergodic rate versus $M_R$ $(=\!\!N_R)$ and $P_I$, respectively, in Special Case (ii) and Special Case (iii). From these figures, we can make the following observations. The analytical rate of the optimal solution $C_{ub}^{(Q)}\left(\boldsymbol\phi^*\right)$ and the rate of the optimal solution $C^{(Q)}\left(\boldsymbol\phi^*\right)$ obtained by Monte Carlo simulation are very close to each other, which verifies that $C_{ub}^{(Q)}(\boldsymbol\phi)$ is a good approximation of $C^{(Q)}(\boldsymbol\phi)$; the rates of the proposed optimal solution and PCD solution are very close in each considered case; the proposed solution in the instantaneous CSI case coincides with the one in \cite{IEEEexample:8746155} in Special Case (ii), and significantly outperforms the one in \cite{IEEEexample:8746155} in Special Case (iii).  From Fig. \ref{fig:optimal_M_R}, we can observe that the rates of the proposed solutions and the design with random phase shifts increase with $M_R$ $(=\!\!N_R)$, mainly due to the increment of reflecting signal power; in Special Case (iii), the average rate of the phase shift design in \cite{IEEEexample:8746155} decreases with $M_R(=N_R)$, revealing the penalty of ignoring interference in phase shift design in the instantaneous CSI case. From Fig. \ref{fig:optimal_P_I}, we can see that the rate of each scheme decreases with $P_I$.
\begin{figure}[t]
\begin{center}
\subfigure[\scriptsize{Instantaneous CSI case.
}\label{fig:3a}]
{\resizebox{4.25cm}{!}{\includegraphics{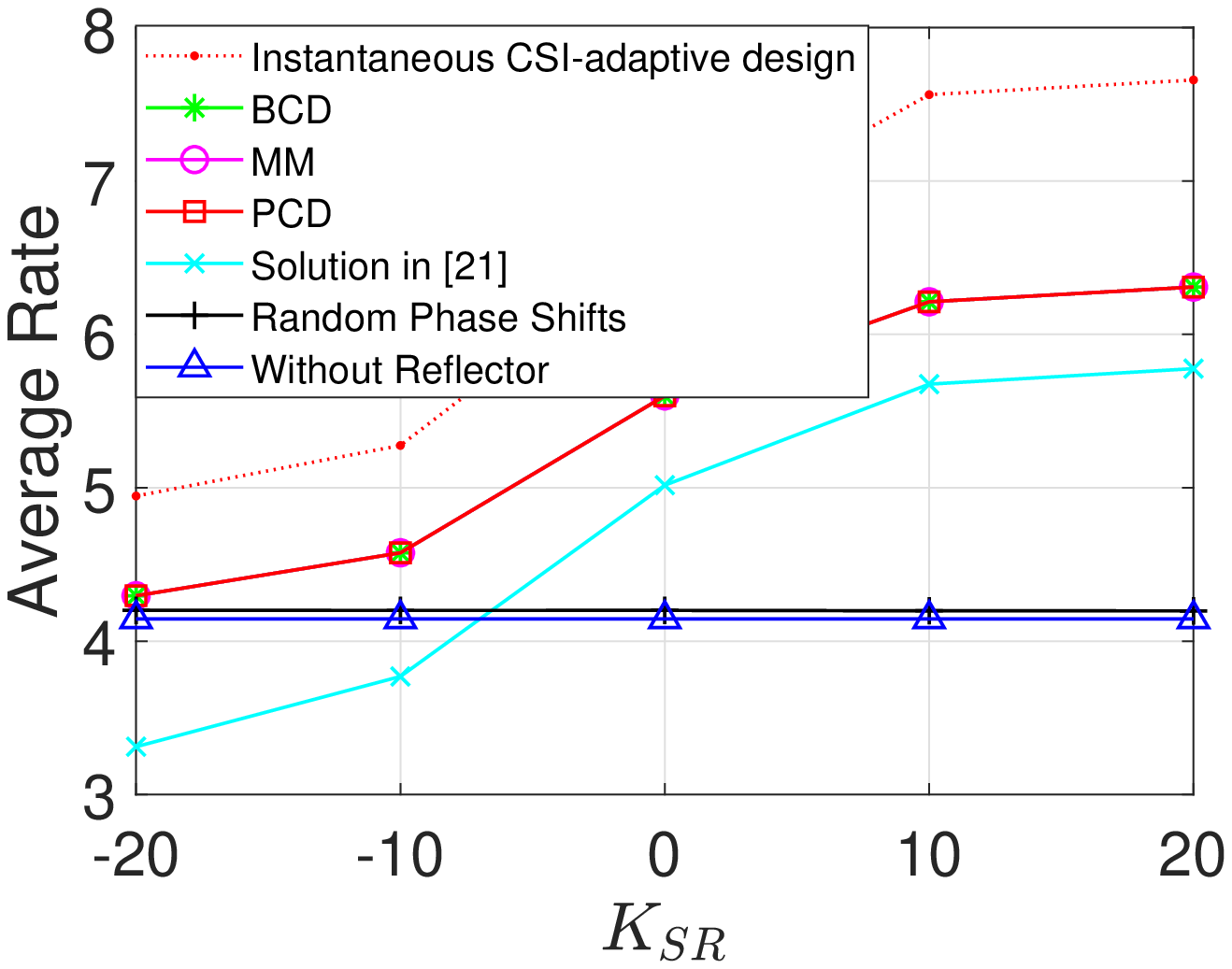}}}\quad
\subfigure[\scriptsize{Statistical CSI case.
}\label{fig:3b}]
{\resizebox{4.25cm}{!}{\includegraphics{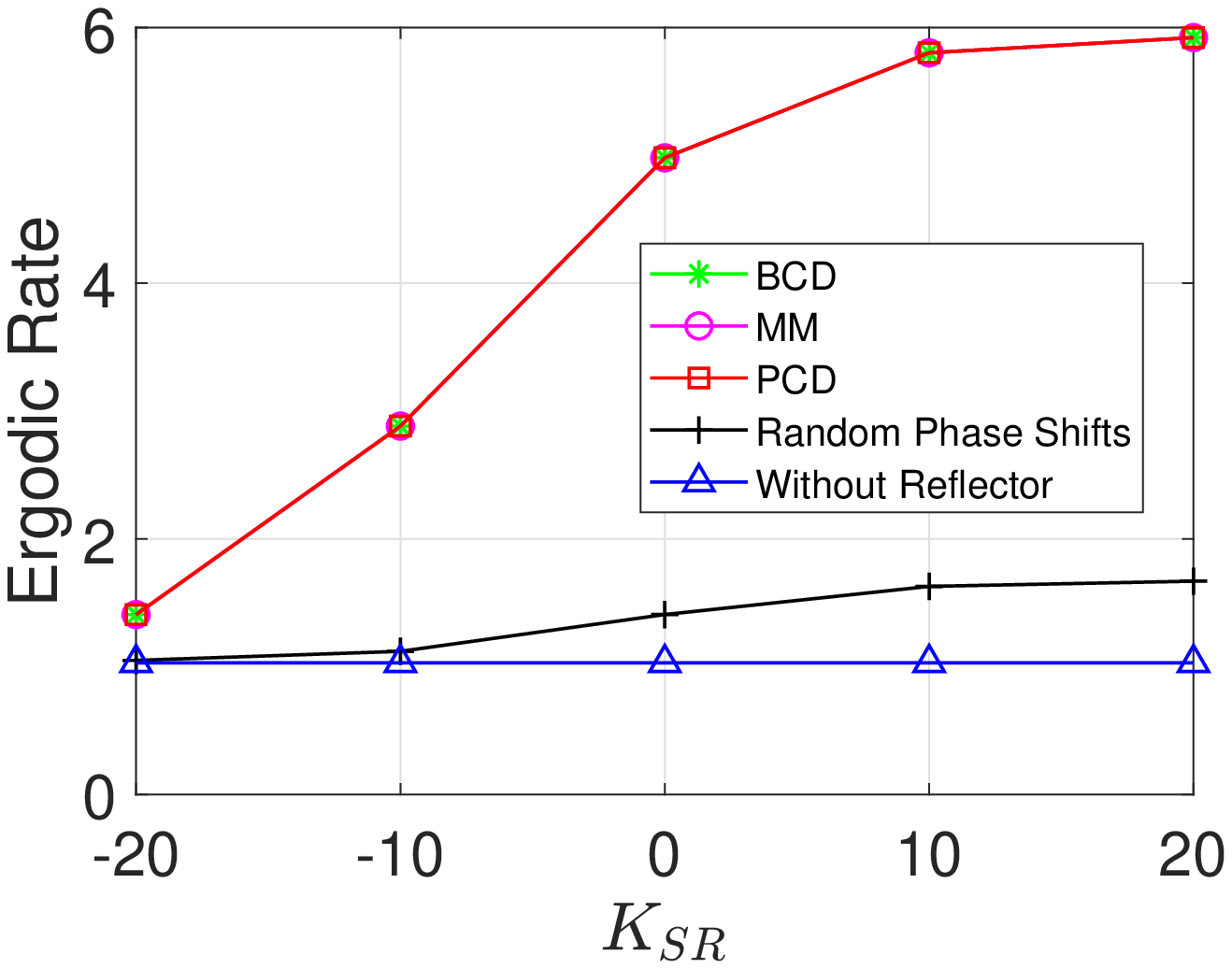}}}
\end{center}
\vspace{-2mm}
\caption{\small{Average rate and ergodic rate versus $K_{SR}$ in the general case.}}
\vspace{-2mm}
\label{fig:General_K_SR}
\end{figure}
\begin{figure}[t]
\begin{center}
\subfigure[\scriptsize{Instantaneous CSI case.
}\label{fig:4a}]
{\resizebox{4.25cm}{!}{\includegraphics{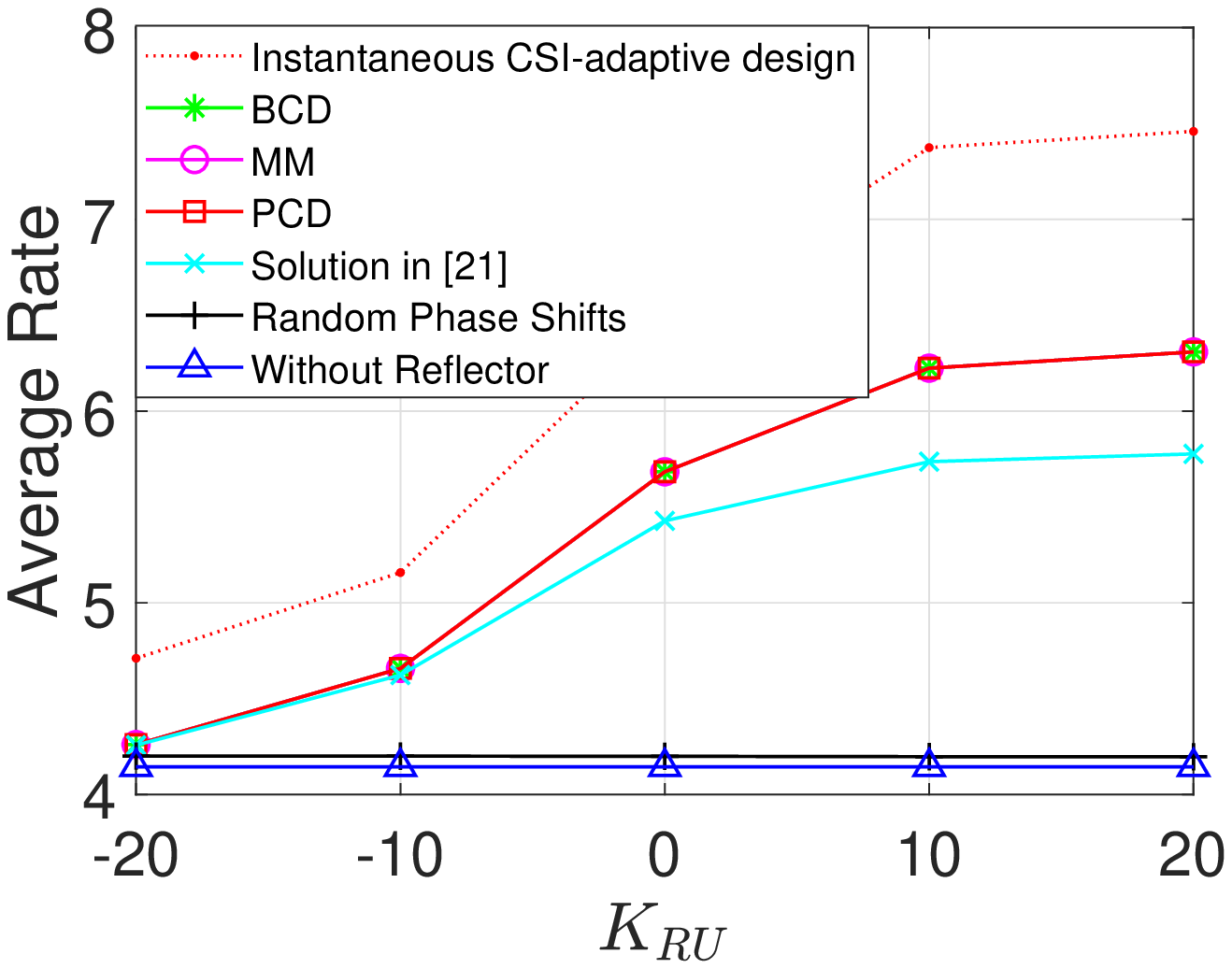}}}\quad
\subfigure[\scriptsize{Statistical CSI case.
}\label{fig:4b}]
{\resizebox{4.25cm}{!}{\includegraphics{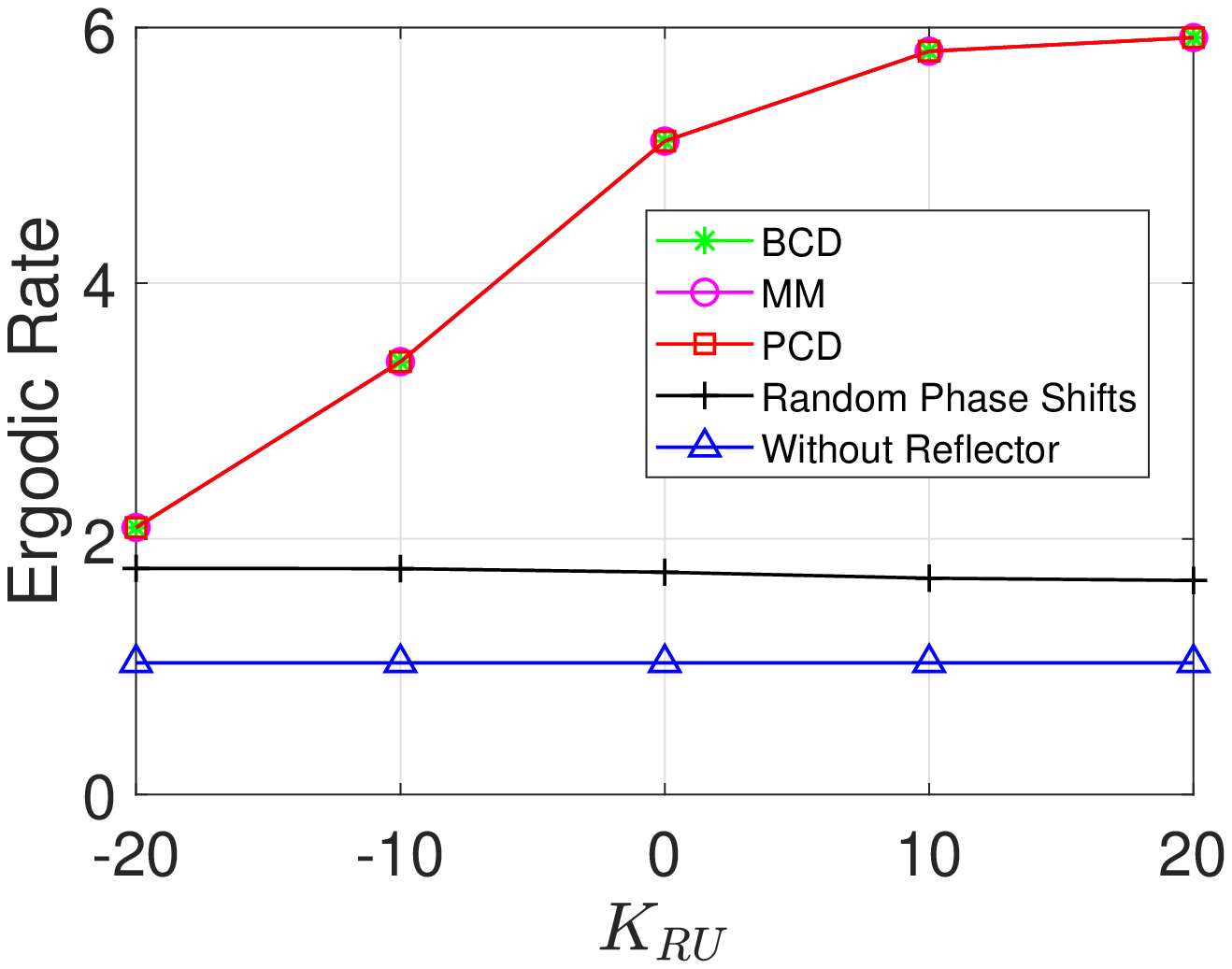}}}
\end{center}
\vspace{-2mm}
\caption{\small{Average rate and ergodic rate versus $K_{RU}$ in the general case.}}
\vspace{-2mm}
\label{fig:General_K_RU}
\end{figure}
\begin{figure}[t]
\begin{center}
\subfigure[\scriptsize{Instantaneous CSI case.
}\label{fig:5a}]
{\resizebox{4cm}{!}{\includegraphics{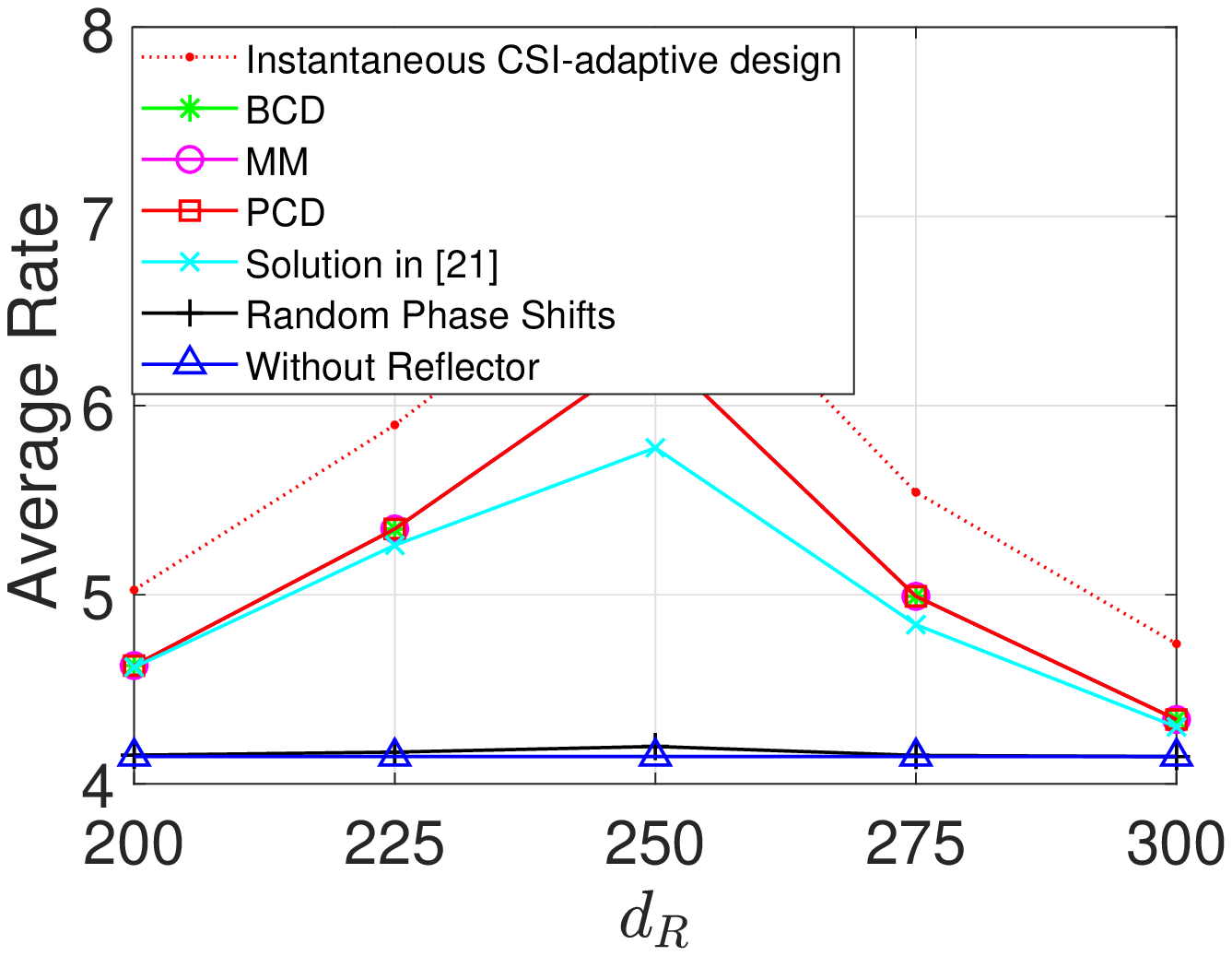}}}\quad
\subfigure[\scriptsize{Statistical CSI case.
}\label{fig:5b}]
{\resizebox{4cm}{!}{\includegraphics{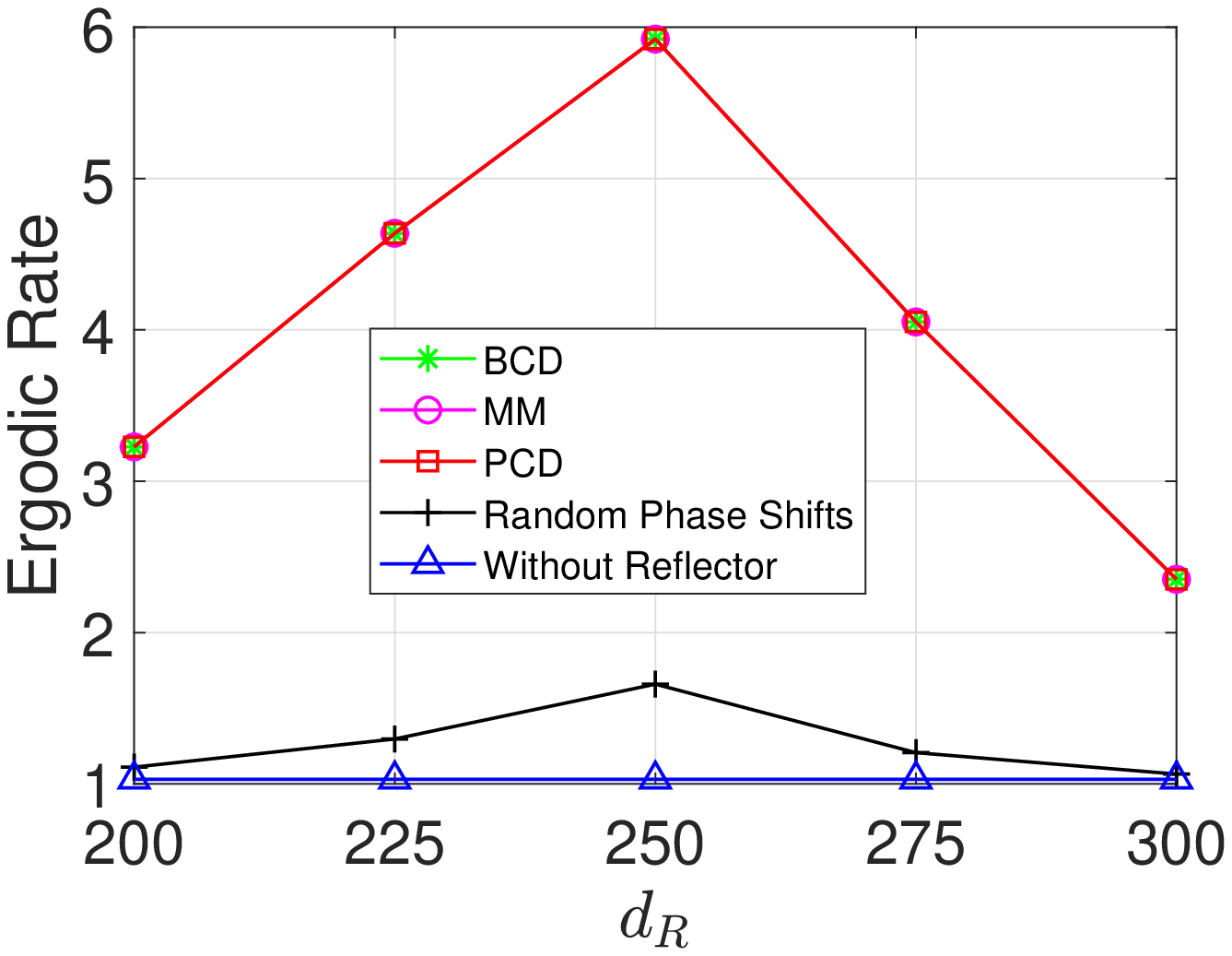}}}
\end{center}
\vspace{-2mm}
\caption{\small{Average rate and ergodic rate versus $d_R$ in the general case.}}
\vspace{-2mm}
\label{fig:General_d_R}
\end{figure}
\begin{figure}[t]
\begin{center}
\subfigure[\scriptsize{Instantaneous CSI case.
}\label{fig:6a}]
{\resizebox{4.25cm}{!}{\includegraphics{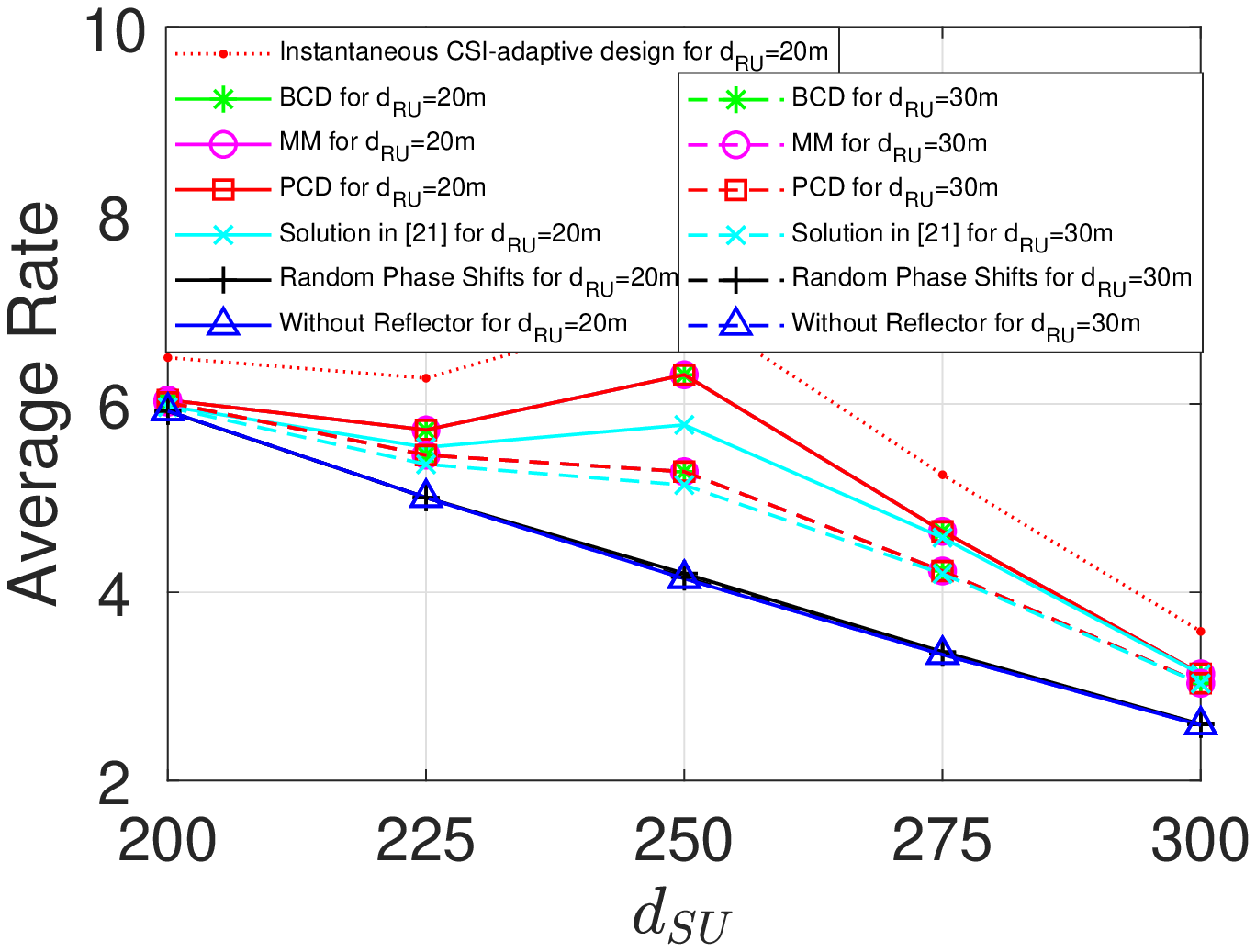}}}\quad
\subfigure[\scriptsize{Statistical CSI case.
}\label{fig:6b}]
{\resizebox{4.25cm}{!}{\includegraphics{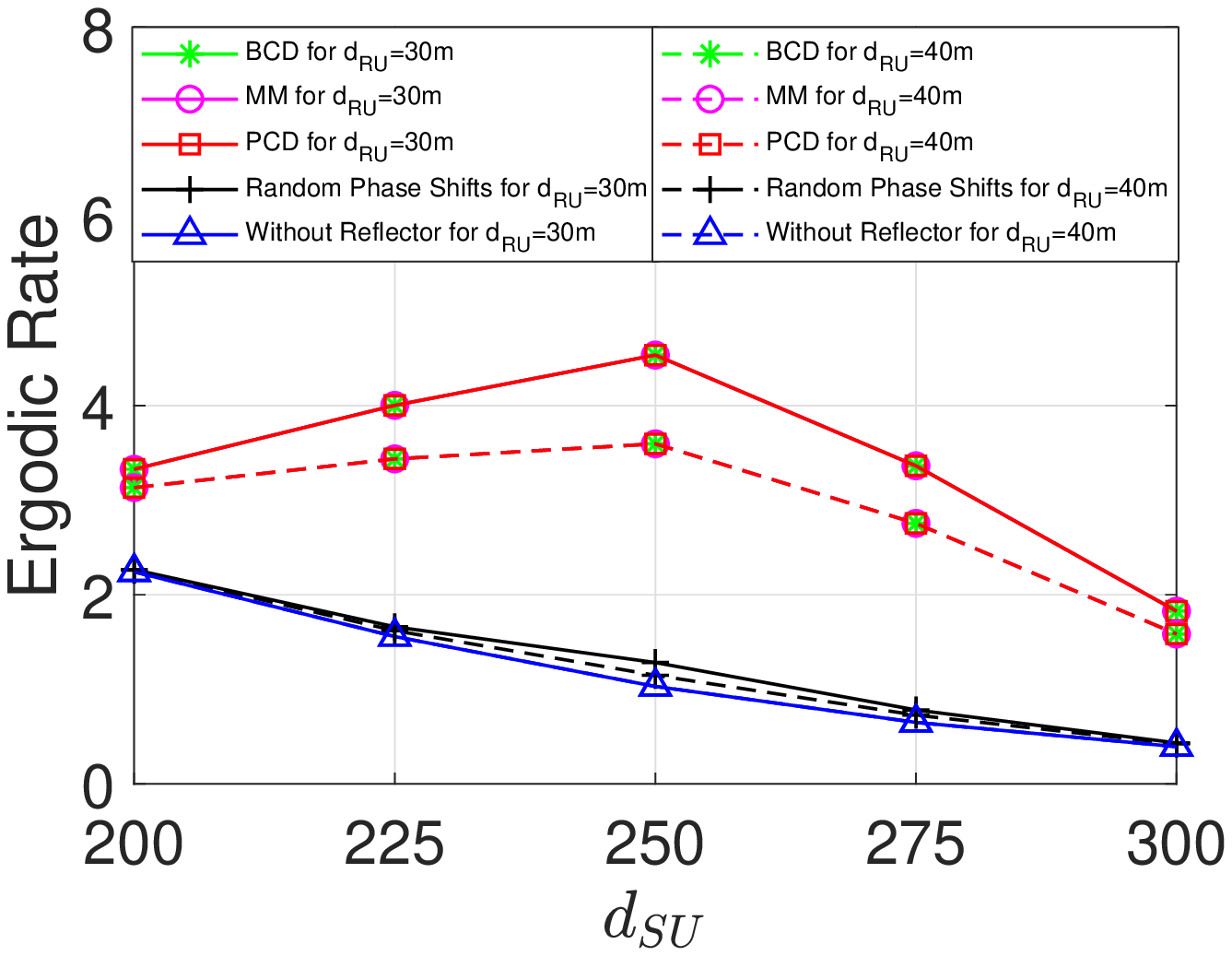}}}
\end{center}
\vspace{-2mm}
\caption{\small{Average rate and ergodic rate versus $d_{SU}$ in the general case.}}
\vspace{-2mm}
\label{fig:General_d_SU}
\end{figure}
\begin{figure}[t]
\begin{center}
\subfigure[\scriptsize{Instantaneous CSI case.
}\label{fig:8a}]
{\resizebox{4.25cm}{!}{\includegraphics{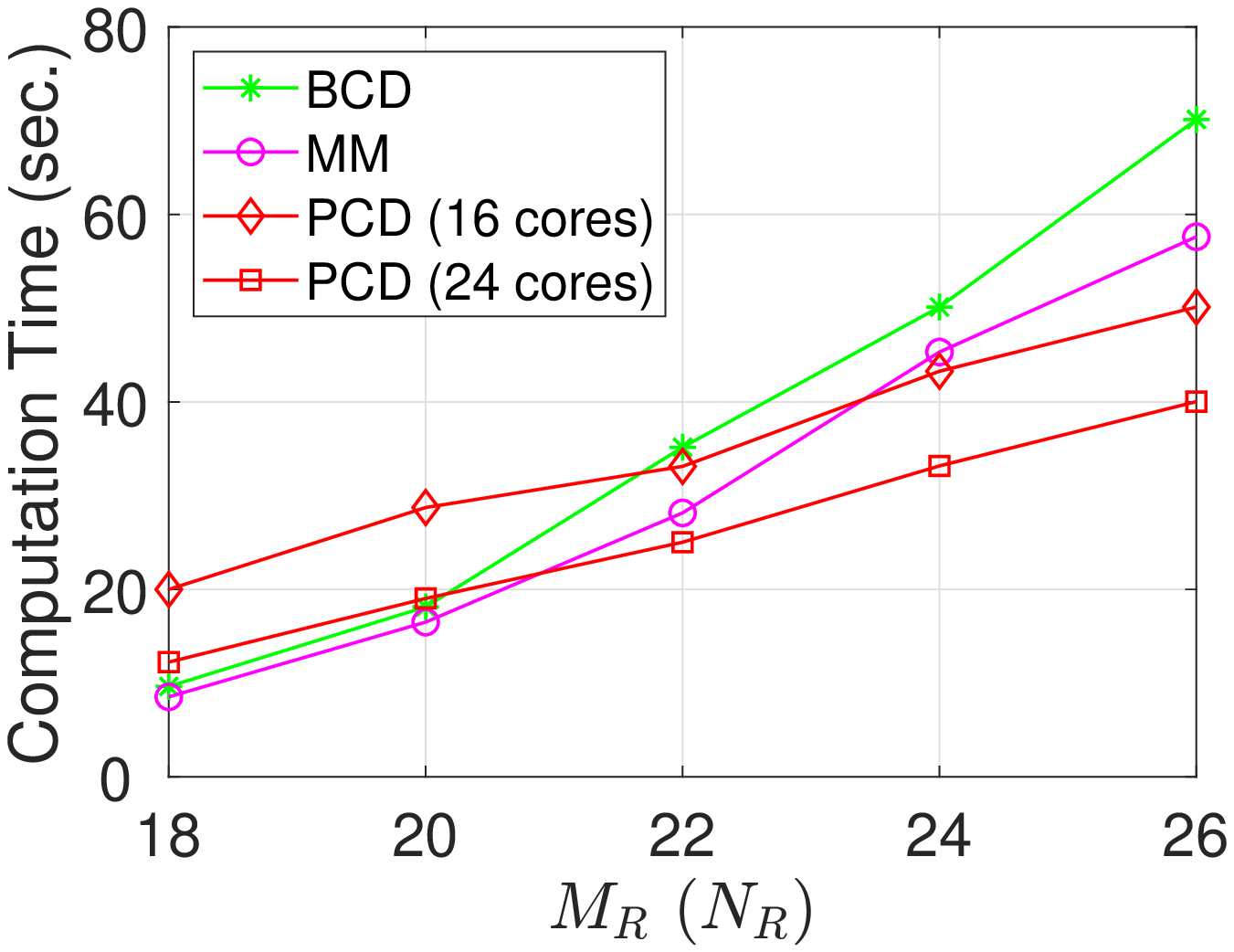}}}\quad
\subfigure[\scriptsize{Statitsical CSI case.
}\label{fig:8b}]
{\resizebox{4.25cm}{!}{\includegraphics{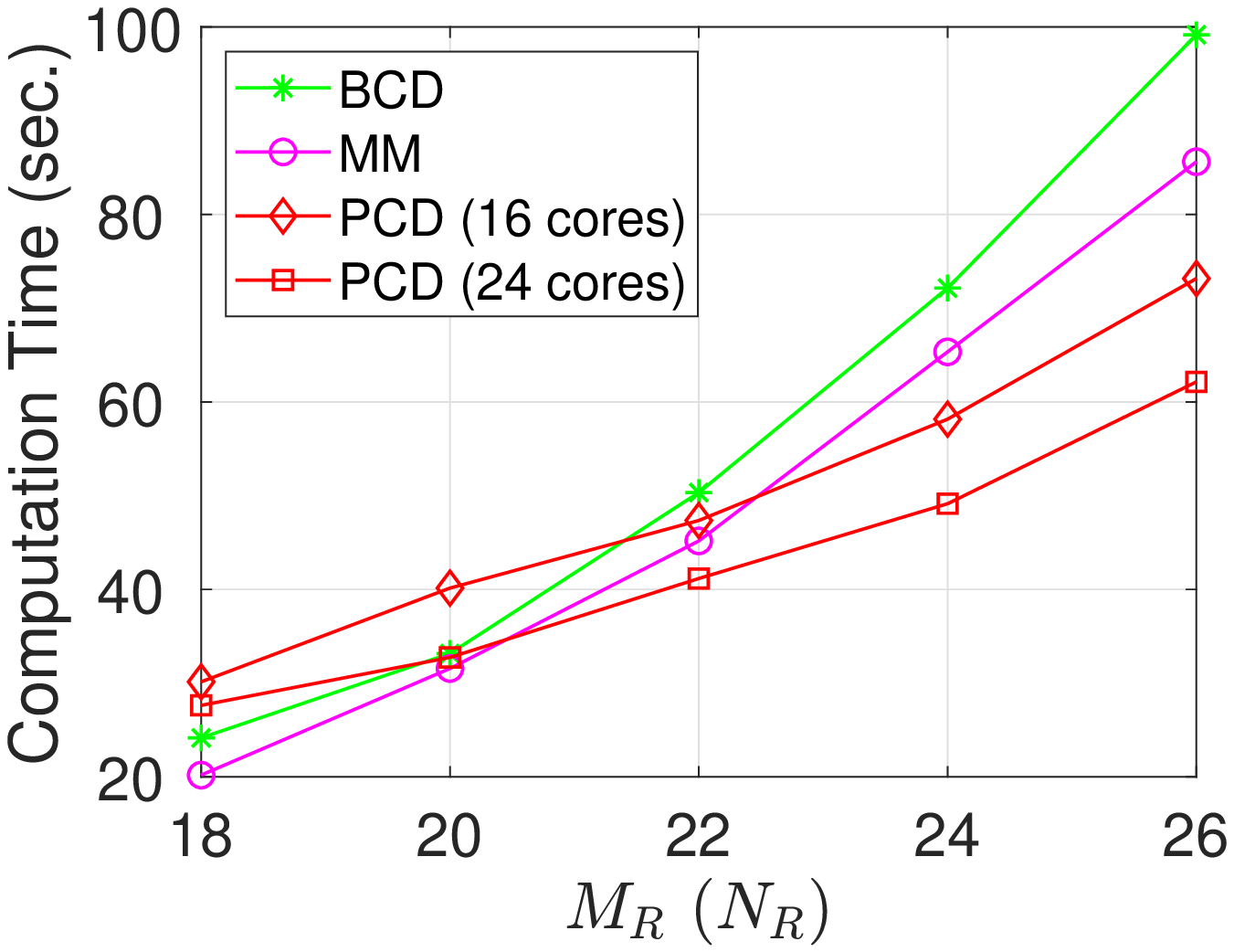}}}
\end{center}
\vspace{-2mm}
\caption{\small{Running time versus $M_R$ $(=N_R)$. }}
\vspace{-2mm}
\label{fig:PCD_runningtime}
\end{figure}

In the general case, we set $\delta^{(h)}_{SR}=\delta^{(v)}_{SR}=\pi/6$, $\delta^{(h)}_{IR}=\delta^{(v)}_{IR}=\pi/8$, $K_{SR}=K_{RU}=20$dB, $K_{IR}=10$dB, if not specified otherwise. Fig. \ref{fig:General_K_SR}, Fig. \ref{fig:General_K_RU}, Fig. \ref{fig:General_d_R} and Fig. \ref{fig:General_d_SU} illustrate the average rate and ergodic rate versus $K_{SR}, K_{RU},d_{R}$ and $d_{SU}$, respectively, in the general case. From these figures, we can see that the PCD solution has the same rate as the BCD and MM solutions in each CSI case; the PCD solution significantly outperforms Baseline 2, Baseline 3 and Baseline 4. From Fig. \ref{fig:General_K_SR} and Fig. \ref{fig:General_K_RU}, we can see that the rate of the PCD solution increases with $K_{SR}$ and $K_{RU}$, due to the increment of the channel power of the each LoS component; the fact that the rate of the proposed PCD solution is greater than the rate of the counterpart system without IRS confirms Theorem~\ref{lem:extreme cases1} to certain extent. From Fig. \ref{fig:General_d_R}, we can observe that the rate of the PCD solution increases with $d_{R}$, due to the decrement of the distance between the IRS and user $U$ when $d_{R}<d_{SU}$, and decreases with $d_R$, due to the increment of the distance between the IRS and user $U$ when $d_{R}>d_{SU}$; the rate of the PCD solution in the case of $d_{R}<d_{SU}$ is greater than that in the case of $d_R>d_{SU}$, at the same distance between the IRS and user $U$, due to smaller path loss between the IRS and the signal BS.
From Fig. \ref{fig:General_d_SU}, we can see that in the case of $d_{RU}=20$m, the rate of the PCD solution increases with $d_{SU}$ when $d_{SU}<d_R$, mainly due to the decrement of $d_{RU}$, and decreases with $d_{SU}$ when $d_{SU}>d_R$, due to the increment of both $d_{SU}$ and $d_{RU}$; in the case of $d_{RU}=30$m, the rate of the PCD solution always decreases with $d_{SU}$, mainly due to the increment of the distance between the signal BS and user $U$.

Furthermore, from Fig.~\ref{fig:optimal_M_R} to Fig.~\ref{fig:General_d_SU}, the following observations can be made. For each scheme, the average rate in the instantaneous CSI case is greater than the ergodic rate in the statistical CSI case, which is in accordance with Corollary~\ref{cor:inssta}. When $K_{SR}, K_{RU}, d_{SU}$ are large and $d_{R}$ is small, i.e., $\tau_{SRU}, \alpha_{IU},\alpha_{SR}$ are large, the proposed solution achieves a higher rate than the system without IRS, confirming Theorem~\ref{lem:extreme cases1} to some extent. Under most system parameters, the proposed solution surpasses the one in \cite{IEEEexample:8746155}, indicating the importance of explicitly taking interference into account in designing IRS-assisted systems.

Fig. \ref{fig:PCD_runningtime} illustrates the computation times of the PCD, BCD and MM algorithms versus $M_R$ $(=N_R)$.\footnote{We use MATLAB R2018a in a Ubuntu 18.04 bionic operating system with an AMD Ryzen 9 3900X 24-core CPU.}  From Fig. \ref{fig:PCD_runningtime}, we can see that when the number of IRS elements is large, the gain of the proposed PCD algorithm in computation time over the BCD and MM algorithms increases with the number of the cores on a server, due to its parallel computation mechanism. Note that in practical systems with multi-core processors, the value of the PCD algorithm  will be prominent, especially for large-scale IRS.
\section{Conclusion}
In this paper, we considered the analysis and optimization of quasi-static phase shift design in an IRS-assisted system in the presence of interference. We modeled signal and interference links via the IRS with Rician fading. We considered the instantaneous CSI case and the statistical CSI case, and applied MRT based on the complete CSI and the CSI of the LoS components, respectively. First, we obtained a tractable expression of the average rate in the instantaneous CSI case and a tractable expression of the ergodic rate in the statistical CSI case. We also provided sufficient conditions for the average rate in the instantaneous CSI case to surpass the ergodic rate in the statistical CSI case, at any phase shifts. Then, we considered the average rate maximization and the ergodic rate maximization, both with respect to the phase shifts, which are non-convex problems. For each non-convex problem, we obtained a globally optimal solution under certain system parameters, and proposed the PCD algorithm to obtain a stationary point under arbitrary system parameters.  Next, we characterized sufficient conditions under which the IRS-assisted system with the optimal quasi-static phase shift design is beneficial, compared to the system without IRS. Finally, by numerical results, we verified analytical results and demonstrated notable gains of the proposed solutions over existing schemes. The results in this paper provide important insights for designing practical IRS-assisted systems.
\section*{Appendix A}\label{proof:w_S_statistic}
For notation simplicity, in Appendix A and Appendix B, denote $\mathbf{g}_S \triangleq \mathbf{h}^H_{RU}\Phi(\boldsymbol\phi) \mathbf{H}_{SR}$, $\mathbf{g}_I\triangleq \mathbf{h}^H_{RU}\Phi(\boldsymbol\phi) \mathbf{H}_{IR}$, $\bar{\mathbf{g}}_S \triangleq \bar{\mathbf{h}}^H_{RU}\Phi(\boldsymbol\phi) \bar{\mathbf{H}}_{SR}$ and $\bar{\mathbf{g}}_I\triangleq \bar{\mathbf{h}}^H_{RU}\Phi(\boldsymbol\phi) \bar{\mathbf{H}}_{IR}$.
To show that $\mathbf{w}_S^{(statistic)}$ maximizes $\frac{\mathbb{E}\left[\left\lvert(\mathbf{g}_S + \mathbf{h} ^H_{SU}) \mathbf{w}_S\right\rvert^2\right]}{\mathbb{E}\left[\left\lvert\left(\mathbf{g}_I + \mathbf{h}^H_{IU}\right) \mathbf{w}_I^{(statistic)}\right\rvert^2\right]+\sigma^2}$ subject to $\left\lvert\left\lvert\mathbf{w}_S\right\rvert\right\rvert_2^2=1$, it is equivalent to show that $\mathbf{w}_S^{(statistic)}$ maximizes $\mathbb{E}\left[\left\lvert\left(\mathbf{g}_S + \mathbf{h}^H_{SU}\right)\mathbf{w}_S\right\rvert^2\right]$ subject to $\left\lvert\left\lvert\mathbf{w}_S\right\rvert\right\rvert_2^2=1$. First, we have:
\setcounter{equation}{40}
\begin{align}
&\mathbb{E}\left[\left\lvert\mathbf{g}_S\mathbf{w}_S\right\rvert^2\right]\nonumber \\
\overset{\text{(a)}}{=}& \frac{K_{SR}\alpha_{SR}\alpha_{RU}}{(K_{SR}+1)(K_{RU}+1)}\left(K_{RU}y_{SRU}\left(\boldsymbol\phi\right)
+M_RN_R\right)\nonumber \\ &\times\left\lvert
\mathbf{a}(\varphi^{(h)}_{SR},\varphi^{(v)}_{cR},M_S,N_S)
\mathbf{w}_S\right\rvert^2+\frac{M_RN_R\alpha_{SR}\alpha_{RU}
}{K_{SR}+1}+\alpha_{SU}  \nonumber \\
\overset{\text{(b)}}{\leq}& \frac{K_{SR}M_SN_S\alpha_{SR}\alpha_{RU}\left(K_{RU}y_{SRU}\left(\boldsymbol\phi\right)+M_RN_R\right)}
{(K_{SR}+1)(K_{RU}+1)}\nonumber \\&+\frac{M_RN_R\alpha_{SR}\alpha_{RU}
}{K_{SR}+1}+\alpha_{SU},\label{eq:cauchy}
\end{align}
where $(a)$ is due to $\lvert\lvert\mathbf{w}_{S}\rvert\rvert_2^2=1$ and $(b)$ is due to the Cauchy-Schwartz inequality $
\left\lvert
\mathbf{a}\left(\varphi^{(h)}_{SR},\varphi^{(v)}_{cR},M_S,N_S\right)
\mathbf{w}_S\right\rvert^2 \leq M_SN_S$.
Note that the equality holds when $\mathbf{w}_S=\frac{{\bar{\mathbf{g}}_S}^H}{\left\lvert\left \lvert \bar{\mathbf{g}}_S \right \rvert \right\rvert_2}e^{j\alpha}$, for all $\alpha \in [0,2\pi)$. By setting $\alpha=0$, we can obtain $\mathbf{w}_S^{(statistic)}$. Thus, we can show that $\mathbf{w}_S^{(statistic)}$ maximizes $\frac{\mathbb{E}\left[\left\lvert(\mathbf{g}_S + \mathbf{h} ^H_{SU}) \mathbf{w}_S\right\rvert^2\right]}{\mathbb{E}\left[\left\lvert\left(\mathbf{g}_I + \mathbf{h}^H_{IU}\right) \mathbf{w}_I^{(statistic)}\right\rvert^2\right]+\sigma^2}$ subject to  $\left\lvert\left\lvert\mathbf{w}_S\right\rvert\right\rvert_2^2=1$.
\section*{Appendix B: Proof of Theorem~\ref{lem:ergodic Case 1 with reflector}}\label{proof:lemma ergodic}
First, consider $Q=instant$. By Jensen's inequality, we have:
\begin{align}
&C^{(instant)}(\phi) \leq  \log_2\left(1+\mathbb{E}\left[\gamma^{(instant)}
\left(\boldsymbol\phi\right)\right]\right) \nonumber \\
= & \log_2 \left(1+ \frac{{P_S}\mathbb{E}\left[{\left\lvert\left\lvert \left(\mathbf{g}_S + \mathbf{h}^H_{SU}\right) \right\rvert\right\rvert}_2^2 \right]}{ {P_I}\mathbb{E} \left[ {\left\lvert \left(\mathbf{g}_I + \mathbf{h}^H_{IU}\right)  \frac{\mathbf{h}_{IU'}}{\left\lvert\left\lvert \mathbf{h}_{IU'} \right\rvert\right\rvert_2} \right\rvert}^2 \right]+ {\sigma}^2}\right). \label{eq:jensen}
\end{align}
We calculate ${ \mathbb{E}\left[ {\left\lvert \left(\mathbf{g}_I + \mathbf{h}^H_{IU}\right)  \frac{\mathbf{h}_{IU'}}{\left\lvert\left\lvert \mathbf{h}_{IU'} \right\rvert\right\rvert_2}\right \rvert}^2 \right]}$ as follows:
\begin{align}
&{ \mathbb{E}\left[ {\left\lvert \left(\mathbf{g}_I + \mathbf{h}^H_{IU}\right)  \frac{\mathbf{h}_{IU'}}{\left\lvert\left\lvert \mathbf{h}_{IU'} \right\rvert\right\rvert_2}\right \rvert}^2 \right]}\nonumber \\
= &{ \mathbb{E}\left [ { \left(\mathbf{g}_I + \mathbf{h}^H_{IU}\right)  \frac{\mathbf{h}_{IU'}\mathbf{h}^H_{IU'}}{\left\lvert\left\lvert \mathbf{h}_{IU'} \right\rvert\right\rvert_2^2} \left(\mathbf{g}_I + \mathbf{h}^H_{IU}\right)^H}\right]}\nonumber \\
= &{\mathbb{E}\left[ { \left(\mathbf{g}_I + \mathbf{h}^H_{IU}\right)  \mathbb{E}\left[\frac{\mathbf{h}_{IU'}\mathbf{h}^H_{IU'}
}{\left\lvert\left\lvert \mathbf{h}_{IU'} \right\rvert\right\rvert_2^2}\right] \left(\mathbf{g}_I+ \mathbf{h}^H_{IU}\right)^H}\right]} \nonumber \\
\overset{\text{(a)}}{=}& \frac{1}{M_IN_I}\left(\mathbb{E}\left[
\left\lvert\left\lvert \mathbf{g}_I\right\rvert\right\rvert^2_2\right]
+2\mathbb{E}\left[\mathbf{g}_I\mathbf{h}_{IU}\right]+\mathbb{E}
\left[\left\lvert\left\lvert\mathbf{h}^H_{IU}
\right\rvert\right\rvert^2_2\right]\right)\nonumber \\
\overset{\text{(b)}}{=}& \alpha_{IR}\alpha_{RU}\!\left(\tau_{IRU}y_{IRU}(\boldsymbol\phi)
\!+\!(1-\tau_{IRU})M_RN_R\right)+\alpha_{IU},\label{eq:ibs}
\end{align}where $(a)$ is due to $ \mathbb{E}\left[\frac{\mathbf{h}_{IU'}
\mathbf{h}^H_{IU'}}{\left\lvert\left\lvert \mathbf{h}_{IU'} \right\rvert\right\rvert_2^2}\right]=$ $\frac{1}{MN}\mathbf{I}_{MN}$ with $\mathbf{I}_{MN}$ representing the $MN\times MN$ identity matrix, and $(b)$ is due to $\mathbb{E}\left[\left\lvert\left\lvert\mathbf{g}_I\right\rvert\right\rvert^2_2\right]= \alpha_{IR}\alpha_{RU}M_IN_I\left(\tau_{IRU}y_{IRU}(\boldsymbol\phi)+(1-\tau_{IRU})\right.$ $\left.M_RN_R\right)$, $\mathbb{E}\left[\mathbf{g}_I\mathbf{h}_{IU}\right]=0$ and $\mathbb{E}\left[\left\lvert\left\lvert\mathbf{h}^H_{IU}\right\rvert\right\rvert^2_2\right]$ $=\alpha_{IU}$.
Similarly, we have
$\mathbb{E}\left[{\left\lvert\left\lvert \left(\mathbf{g}_S + \mathbf{h}^H_{SU}\right) \right\rvert\right\rvert}_2^2 \right]
= M_SN_S\alpha_{SR}\alpha_{RU}\left(\tau_{SRU}y_{SRU}(\boldsymbol\phi)+(1-\tau_{SRU})M_RN_R\right) +M_SN_S\alpha_{SU}$.
Thus, we have $ C^{(instant)}(\boldsymbol\phi) \leq C^{(instant)}_{ub}(\boldsymbol\phi)$.

Next, consider $Q=statistic$.
Similarly, by Jensen's inequality, we have
$C^{(statistic)}(\boldsymbol\phi) \leq \log_2\left(1 + {  \frac{\mathbb{E}\left[{P_S}{\left\lvert   \frac{\left(\mathbf{g}_S + \mathbf{h}^H_{SU} \right){\bar{\mathbf{g}}_S}^H}{\left\lvert\left\lvert \bar{\mathbf{g}}_S  \right\rvert\right\rvert_2}\right  \rvert}^2\right]}{ {P_I}\mathbb{E} \left[ {\left\lvert \left(\mathbf{g}_I + \mathbf{h}^H_{IU}\right)\frac{\boldsymbol{1}_{M_IN_I}}{\sqrt{M_IN_I}} \right\rvert}^2 \right]+ {\sigma}^2}}\right)$.
We calculate $\mathbb{E}\left[{\left\lvert  \frac{\left(\mathbf{g}_S + \mathbf{h}^H_{SU}\right){\bar{\mathbf{g}}_S}^H}{\left\lvert\left\lvert \bar{\mathbf{g}}_S  \right\rvert\right\rvert_2}\right  \rvert}^2\right]$ as follows:
\begin{align*}
\mathbb{E}\left[{\left\lvert   \frac{\left(\mathbf{g}_S + \mathbf{h}^H_{SU} \right) {\bar{\mathbf{g}}_S}^H}{\left\lvert\left\lvert \bar{\mathbf{g}}_S  \right\rvert\right\rvert_2}\right  \rvert}^2\right]
= \mathbb{E}\left[{\left\lvert  \frac{ \mathbf{g}_S{\bar{\mathbf{g}}_S}^H}
{\left\lvert\left\lvert \bar{\mathbf{g}}_S  \right\rvert\right\rvert_2}\right  \rvert}^2\right]+
\mathbb{E}\left[{\left\lvert   \frac{\mathbf{h}^H_{SU} \bar{\mathbf{g}}_S^H}{\left\lvert\left\lvert \bar{\mathbf{g}}_S  \right\rvert\right\rvert_2}\right  \rvert}^2\right]  \end{align*}
\begin{align}
\overset{\text{(c)}}{=}&{M_SN_S}\left( \alpha_{SR}\alpha_{RU}\left(\tau_{SRU} y_{SRU}(\boldsymbol\phi) + \left(1-\tau_{SRU}\right.\right. \right.\nonumber \\
&\left. \left. \left. -\frac{M_SN_S-1}{M_SN_S(K_{SR}+1)}\right)M_RN_R \right)\right)+\alpha_{SU}, \label{eq:lemma2_1}
\end{align}
where $(c)$ is due to $ \mathbb{E}\left[{\left\lvert   \frac{\mathbf{g}_S {\bar{\mathbf{g}}_S}^H}{\left\lvert\left\lvert \bar{\mathbf{g}}_S  \right\rvert\right\rvert_2}\right  \rvert}^2\right]=M_SN_S\alpha_{SR}\alpha_{RU}
$ $\left(\tau_{SRU} y_{SRU}(\boldsymbol\phi) + \left(1-\tau_{SRU}-\frac{M_SN_S-1}{M_SN_S(K_{SR}+1)}\right)M_RN_R \right)$ and $\mathbb{E}\left[{\left\lvert  \frac{\mathbf{h}^H_{SU} {\bar{\mathbf{g}}_S}^H}{\left\lvert\left\lvert \bar{\mathbf{g}}_S \right\rvert\right\rvert_2}\right  \rvert}^2\right]=\alpha_{SU}$. Similarly, we have
$
\mathbb{E} \left[ {\left\lvert \left(\mathbf{g}_I + \mathbf{h}^H_{IU}\right)\frac{\boldsymbol{1}_{M_IN_I}}
{\sqrt{M_IN_I}}\right\rvert }^2 \right]
=\alpha_{IR}\alpha_{RU} \left(\frac{y_{IR}\tau_{IRU}}{M_IN_I} y_{IRU}(\boldsymbol\phi)\right.$ $\left. + \left(1-\tau_{IRU} +\frac{\tau_{IRU}(y_{IR}-M_IN_I)}{M_IN_IK_{RU}}\right) M_RN_R\right)+\alpha_{IU}
$.
Thus, we have $ C^{(statistic)}(\boldsymbol\phi) \leq C^{(statistic)}_{ub}(\boldsymbol\phi)$.
\section*{Appendix C: Proof of Theorem~\ref{lem:op}}\label{proof:op}
 First, we consider Special Case (i). For $Q=instant$ or $statistic$, when $M_R=N_R=1$, $y_{SRU}(\boldsymbol\phi)=y_{IRU}(\boldsymbol\phi)=1$ for all $\boldsymbol\phi$. Thus, we can show the statement for Special Case (i).

Next, we consider Special Case (ii) and Special Case (iii). As $\delta^{(h)}_{SR}=\delta^{(h)}_{IR},\delta^{(v)}_{SR}=\delta^{(v)}_{IR}$, we have
$y_{SRU}(\boldsymbol\phi) = y_{IRU}(\boldsymbol\phi) \triangleq y(\boldsymbol\phi)$,
where $y_{SRU}(\boldsymbol\phi)$ and $y_{IRU}(\boldsymbol\phi)$ are given by \eqref{eq:over gamma}. Thus, by \eqref{eq:ub gamma}, we have
$
\gamma^{(Q)}_{ub}(\boldsymbol\phi) =
 \frac{A_{SRU,LoS} y(\boldsymbol\phi) + A_{SRU,NLoS}^{(Q)}+A_{SU}^{(Q)} }{A_{IRU,LoS}^{(Q)}y(\boldsymbol\phi) + A_{IRU,NLoS}^{(Q)}+A_{IU}} \triangleq\tilde{\gamma}^{(Q)}_{ub}(y(\boldsymbol\phi)),
$
i.e., $\gamma^{(Q)}_{ub}=\tilde{\gamma}^{(Q)}_{ub} \circ y $, where $\circ$ denotes the function composition. The derivative of $\tilde{\gamma}^{(Q)}_{ub}$ is given by:
\begin{align*}
\frac{\mathrm{d}{\tilde{\gamma}^{(Q)}_{ub}}}{\mathrm{d}y}=
\frac{\eta^{(Q)}}
{\left( A_{IRU,LoS}^{(Q)}y(\boldsymbol\phi) + A_{IRU,NLoS}^{(Q)}+A_{IU}\right)^2}.
\end{align*}
\begin{itemize}
\item Consider Special Case (ii). For $Q=instant$ or $statistic$, $\eta^{(Q)}>0$ implies $\frac{\mathrm{d}{\tilde{\gamma}^{(Q)}_{ub}}}{\mathrm{d}y}>0$. Thus, Problem \ref{prob:eq} is equivalent to the following problem:
\begin{align*}
&\max_{\boldsymbol\phi}\ y(\boldsymbol{\phi})\\
&s.t.\quad \eqref{eq:phi}.
\end{align*}
By the triangle inequality, we have:
\begin{align*}
y(\boldsymbol{\phi})\leq \left(\sum\limits_{m=1}^{M_R}\sum\limits_{n=1}^{N_R}{\left\lvert e^{j\theta_{IRU,m,n}+j\phi_{m,n}}\right\rvert}\right)^2=M_R^2N^2_R,
\end{align*}
where the equality holds when $\theta_{IRU,m,n}+\phi_{m,n}=\alpha,m \in \mathcal M_R, n \in \mathcal N_R$, for all $\alpha \in \mathbb{R}$. Thus, we can show the statement for Special Case (ii).
\item Consider Special Case (iii). For $Q=instant$ or $Q=statistic$, $\eta^{(Q)}<0$ implies $\frac{\mathrm{d}{\tilde{\gamma}^{(Q)}_{ub}}}{\mathrm{d}y}\leq0$. Thus, Problem~\ref{prob:eq} is equivalent to the following problem:
\begin{align*}
&\min_{\boldsymbol\phi}\ y(\boldsymbol{\phi})\\
&s.t.\quad \eqref{eq:phi}.
\end{align*} $y(\boldsymbol{\phi})={\left\lvert\sum\limits_{m=1}^{M_R}\sum\limits_{n=1}^{N_R} e^{j\theta_{IRU,m,n}+j\phi_{m,n}}\right\rvert}^2\geq 0$, where the equality holds when $e^{j\left(\theta_{IRU,m,2i-1}+\phi_{m,2i-1}\right)}$ $+e^{j\left(\theta_{IRU,m,2i}+\phi_{m,2i}\right)}=0, m \in \mathcal M_R, i=1,...,\frac{N_R}{2}$, i.e., $\phi_{m,2i}- \phi_{m,2i-1}=(2k_i+1)\pi-\left(\theta_{IRU,m,2i}-\theta_{IRU,m,2i-1}\right)$ for some $k_i \in \mathbb{Z}, m \in \mathcal M_R, i=1,...,\frac{N_R}{2}$.  Thus, we can show the statement for Special Case (iii).
\end{itemize}

Finally, we consider Special Case (iv). From the proof of the statement for Special Case (ii), we can easily show the statement for Special Case (iv).
\begin{figure*}
\begin{align*}
&C_{ub}^{(Q)}(\boldsymbol\phi^\dag)-C^{(Q)}_{ub}
(\boldsymbol\phi^\dag+\boldsymbol\delta)
=\sum\limits_{k=1}^{M_R}
\sum\limits_{l=1}^{N_R}
-\delta_{k,l}\frac{\partial C_{ub}^{(Q)}\left(\overline{\phi}_{k,l}\right)}{\partial \overline{\phi}_{k,l}}
  \leq \sum\limits_{k=1}^{M_R}
\sum\limits_{l=1}^{N_R}
\left\lvert\delta_{k,l}\right\rvert \left\lvert\frac{\partial C_{ub}^{(Q)}\left(\overline{\phi}_{k,l}\right)}{\partial \overline{\phi}_{k,l}}\right\rvert \\
\overset{(c)}{\leq} & \frac{2\pi M_RN_R\left\lvert A_{IRU,LoS}^{(Q)}\left(A_{SRU,NLoS}^{(Q)}+
A_{SU}^{(Q)}\right)-A_{SRU,LoS}\left(
A_{IRU,NLoS}^{(Q)}+A_{IU}\right)\right\rvert(M_RN_R-1)}
{2^b\ln2\left(A_{IRU,NLoS}^{(Q)}+A_{IU}\right)
\left(A_{SRU,NLoS}^{(Q)}+A_{SU}^{(Q)}+A_{IRU,NLoS}^{(Q)}+A_{IU}\right)}
\end{align*}
\hrulefill
\vspace{-1mm}
\end{figure*}\section*{Appendix D: Proof of Theorem~\ref{lem:quantize}}\label{proof:quantize}
First, we consider the special cases. By Theorem~\ref{lem:op}, we have $y_{SRU}\left(\boldsymbol\phi\right)=
y_{IRU}\left(\boldsymbol\phi\right)=
y\left(\boldsymbol\phi\right)=1$ for all $\boldsymbol\phi$ in Special Case (i). Thus, in Special Case (i), $y\left(\boldsymbol\phi^*\right)
=y\left(\boldsymbol\phi^*+\boldsymbol\delta\right)$, implying $\zeta^{(Q)}\left(\boldsymbol\phi^*\right)=0$.
In addition, by Theorem~\ref{lem:op},
\begin{align}
&\left(y_{SRU}\left(\boldsymbol\phi^*\right),
y_{IRU}\left(\boldsymbol\phi^*\right)\right)\nonumber \\ =&
\begin{cases}
\left(M_R^2N_R^2,M_R^2N_R^2\right), & \text{Special Case (ii)},\\
(0,0), & \text{Special Case (iii)} , \\
\left(M_R^2N_R^2,0\right), & \text{Special Case (iv)} ,
\end{cases}\label{eq:gammathree}
\end{align}
implying
\begin{align}
C_{ub}^{(Q)}(\boldsymbol\phi^*)=\!
\begin{cases}\!
\log_2\!\left(1\!+\!\frac{A_{SRU,LoS}M_R^2N_R^2+A_{SRU,NLoS}^{(Q)}
+A_{SU}^{(Q)}}{A_{IRU,LoS}^{(Q)}
M_R^2N_R^2+A_{IRU,NLoS}^{(Q)}+A_{IU}}\right),\\
\log_2\!\left(1\!+\!\frac{A_{SRU,NLoS}^{(Q)}
+A_{SU}^{(Q)}}{A_{IRU,NLoS}^{(Q)}+A_{IU}}\right), \\
\log_2\!\left(1\!+\!\frac{A_{SRU,LoS}M_R^2N_R^2+A_{SRU,NLoS}^{(Q)}
+A_{SU}^{(Q)}}{\sigma^2}\right).
\end{cases}\label{eq:ratethree}
\end{align}
From the proof for Theorem~\ref{lem:op},  $y_{SRU}\left(\boldsymbol\phi\right)=
y_{IRU}\left(\boldsymbol\phi\right)=
y\left(\boldsymbol\phi\right)$ for all $\boldsymbol\phi$ in Special Cases (ii) and (iii), and $\gamma_{ub}^{(Q)}\left(\boldsymbol\phi\right)$ increases with $y\left(\boldsymbol\phi\right)$ in Special Case (ii) and decreases with $y\left(\boldsymbol\phi\right)$ in  Special Case (iii). Then, in Special Case (ii),
$
C^{(Q)}_{ub}
(\boldsymbol\phi^*+\boldsymbol\delta)
\geq  \log_2\left(1+\frac{4\left\lceil \frac{M_RN_R-1}{2}\right \rceil^2A_{SRU,LoS}
\cos^2\frac{2\pi}{2^{b+1}}+A_{SRU,NLoS}^{(Q)}
+A_{SU}^{(Q)}}{4\left\lceil \frac{M_RN_R-1}{2} \right\rceil^2A_{IRU,LoS}^{(Q)}
\cos^2\frac{2\pi}{2^{b+1}}+A_{IRU,NLoS}^{(Q)}+A_{IU}}\right)
$,
where the inequality is due to that $\gamma_{ub}^{(Q)}\left(\boldsymbol\phi\right)$ increases with $y\left(\boldsymbol\phi\right)$ and
$
\left\lvert \sum\limits_{n=1}^{N_R}\sum\limits_{m=1}^{M_R}e^{j\delta_{m,n}}\right\rvert^2 \geq \left\lvert\left\lceil \frac{M_RN_R-1}{2} \right\rceil\left(e^{j\frac{2\pi}{b+1}}+
e^{-j\frac{2\pi}{2^{b+1}}}\right)\right\rvert^2\!=\!
4\left\lceil \frac{M_RN_R-1}{2} \right\rceil^2\cos^2\left(\frac{\pi}{2^b}\right)
$.
In Special Case (iii),
$
C^{(Q)}_{ub}
(\boldsymbol\phi^*+\boldsymbol\delta)
\geq  \log_2\left(1+\frac{4\left\lceil \frac{M_RN_R-1}{2}\right \rceil^2A_{SRU,LoS}
\sin^2\frac{2\pi}{2^{b+1}}+A_{SRU,NLoS}^{(Q)}
+A_{SU}^{(Q)}}{4\left\lceil \frac{M_RN_R-1}{2} \right\rceil^2A_{IRU,LoS}^{(Q)}
\sin^2\frac{2\pi}{2^{b+1}}+A_{IRU,NLoS}^{(Q)}+A_{IU}}\right)
$,
where the inequality is due to that $\gamma_{ub}^{(Q)}\left(\boldsymbol\phi\right)$ decreases with $y\left(\boldsymbol\phi\right)$ and
$\left\lvert\sum\limits_{n=1}^{N_R}\sum\limits_{m=1}^{M_R}
(-1)^{m+n}e^{j\delta_{m,n}}\right\rvert^2
 \leq \left\lvert\left\lceil \frac{M_RN_R-1}{2} \right\rceil\left(e^{j\frac{2\pi}{b+1}}-
e^{-j\frac{2\pi}{2^{b+1}}}\right)\right\rvert^2
\!=\!4\left\lceil \frac{M_RN_R-1}{2} \right\rceil^2\!\sin^2\left(\frac{\pi}{2^b}\right)
$.
In Special Case (iv), by \eqref{eq:withoutinterferenceBS},
$
C^{(Q)}_{ub}
(\boldsymbol\phi^*+\boldsymbol\delta)
\geq  \log_2\left(1+\frac{4\left\lceil \frac{M_RN_R-1}{2}\right \rceil^2A_{SRU,LoS}
\sin^2\frac{2\pi}{2^{b+1}}+A_{SRU,NLoS}^{(Q)}
+A_{SU}^{(Q)}}{\sigma^2}\right)
$.
Thus, by \eqref{eq:ratethree}, we can show \eqref{eq:q1}, \eqref{eq:q2} and \eqref{eq:q4}.

Next, we consider the general case. By mean value theorem, $\zeta^{(Q)}\left(\boldsymbol\phi^\dag\right)$ can be upper bounded as shown at the top of the page,
where $\overline{\phi}_{k,l}$ is between $\phi^\dag_{k,l}$ and $\phi^\dag_{k,l}+\delta_{k,l}$ for all $k \in \mathcal M_R, l \in \mathcal N_R$, and (c) is due to $\left\lvert\delta_{k,l}\right\rvert \leq \frac{2\pi}{2^{b+1}}$.

Finally, we show the monotonicity of each upper bound. It is obvious that the upper bounds in \eqref{eq:q4} and \eqref{eq:q3} decrease with $b$. We know that $\frac{4\left\lceil \frac{M_RN_R-1}{2}\right \rceil^2A_{SRU,LoS}y+A_{SRU,NLoS}^{(Q)}
+A_{SU}^{(Q)}}{4\left\lceil \frac{M_RN_R-1}{2}\right \rceil^2A_{IRU,LoS}^{(Q)}y+A_{IRU,NLoS}^{(Q)}
+A_{IU}}$ increases with $y$ in Special Case (ii) and decreases with $y$ in Special Case (iii). Thus, the upper bounds in \eqref{eq:q1} and \eqref{eq:q2} decrease with $b$.
\section*{Appendix E: Proof of Theorem~\ref{lem:extreme cases1}}\label{proof:comparision1}
First, we consider $\xi^{(Q)}_{>} > 0$. By \eqref{eq:ub gamma},
\begin{align}
&\gamma^{(Q)}_{ub}(\boldsymbol\phi^{*})-\gamma^{(Q)}_{no,ub}
 \overset{\text{(a)}}{\geq}
\gamma^{(Q)}_{ub}(\widetilde{\boldsymbol\phi})-\gamma^{(Q)}_{no,ub}
\nonumber \\\overset{\text{(b)}}{=} & \frac{A_{SRU,LoS}M_R^2N_R^2
+A_{SRU,NLoS}^{(Q)}+A_{SU}^{(Q)}}
{A_{IRU,LoS}^{(Q)}y_I\left(\widetilde{\boldsymbol\phi}\right)+A_{IRU,NLoS}^{(Q)}+A_{IU}}-
\frac{A_{SU}^{(Q)}}{A_{IU}} \nonumber \\
\overset{\text{(c)}}{\geq}& \frac{A_{SRU,LoS}M_R^2N_R^2
+A_{SRU,NLoS}^{(Q)}+A_{SU}^{(Q)}}
{A_{IRU,LoS}^{(Q)}M_R^2N_R^2+A_{IRU,NLoS}^{(Q)}+A_{IU}}-
\frac{A_{SU}^{(Q)}}{A_{IU}} \nonumber \\
= & \frac{\zeta^{(Q)}_{>}}{\left(A_{IRU,LoS}^{(Q)}M_R^2N_R^2+A_{IRU,NLoS}^{(Q)}+A_{IU}
\right)A_{IU}},\label{eq:proof41}
\end{align}
where $\widetilde{\boldsymbol\phi}=\Lambda\left(\alpha-\theta_{SRU,m,n}\right), m \in \mathcal M_R, n \in \mathcal N_R$ for all $\alpha \in \mathbb{R}$, $(a)$ is due to the optimality of $\boldsymbol\phi^*$,  $(b)$ is due to $y_{SRU}\left(\widetilde
{\boldsymbol\phi}\right)=M_R^2N_R^2
$, and $(c)$ is due to $y_{IRU}\left(\boldsymbol\phi\right) \leq M_R^2N_R^2$ for all $\boldsymbol\phi$. By \eqref{eq:proof41}, we know $\zeta^{(Q)}_{>} >0$ implies $\gamma^{(Q)}_{ub}(\boldsymbol\phi^{*})>\gamma^{(Q)}_{no,ub}$.
Next, we consider $\xi^{(Q)}_{<} < 0$. By \eqref{eq:ub gamma},
\begin{align}
&\gamma^{(Q)}_{ub}(\boldsymbol\phi^{*})-\gamma^{(Q)}_{no,ub}
\nonumber\\ \overset{\text{(d)}}{\leq}&
 \frac{A_{SRU,LoS}M_R^2N_R^2+A_{SRU,NLoS}^{(Q)}
+A_{SU}^{(Q)}}{A_{IRU,LoS}^{(Q)}y_{IRU}\left(\boldsymbol\phi^*\right)+
A_{IRU,NLoS}^{(Q)}+A_{IU}}-\frac{A_{SU}^{(Q)}}{A_{IU}}
\nonumber\\
\overset{\text{(e)}}{\leq}
& \frac{A_{SRU,LoS}M_R^2N_R^2+A_{SRU,NLoS}^{(Q)}
+A_{SU}^{(Q)}}{A_{IRU,NLoS}^{(Q)}+A_{IU}}-\frac{A_{SU}^{(Q)}}{A_{IU}}
\nonumber\\ = &\frac{\zeta^{(Q)}_{<}}
{\left(A_{IRU,NLoS}^{(Q)}+A_{IU}\right)A_{IU}},
\label{eq:proof42}
\end{align}
where $(d)$ is due to $y_{SRU}(\boldsymbol\phi^*)\leq M_R^2N_R^2$ (by Theorem~\ref{lem:op}), and $(e)$ is due to $y_{IRU}\left(\boldsymbol\phi\right)\geq 0$ for all $\boldsymbol\phi$.
By \eqref{eq:proof42}, we know $\zeta^{(Q)}_{<}<0$ implies $\gamma^{(Q)}_{ub}(\boldsymbol\phi^{*})<\gamma^{(Q)}_{no,ub}$.




\bibliographystyle{IEEEtran}

\begin{IEEEbiography}[{\includegraphics[height=1.25in]{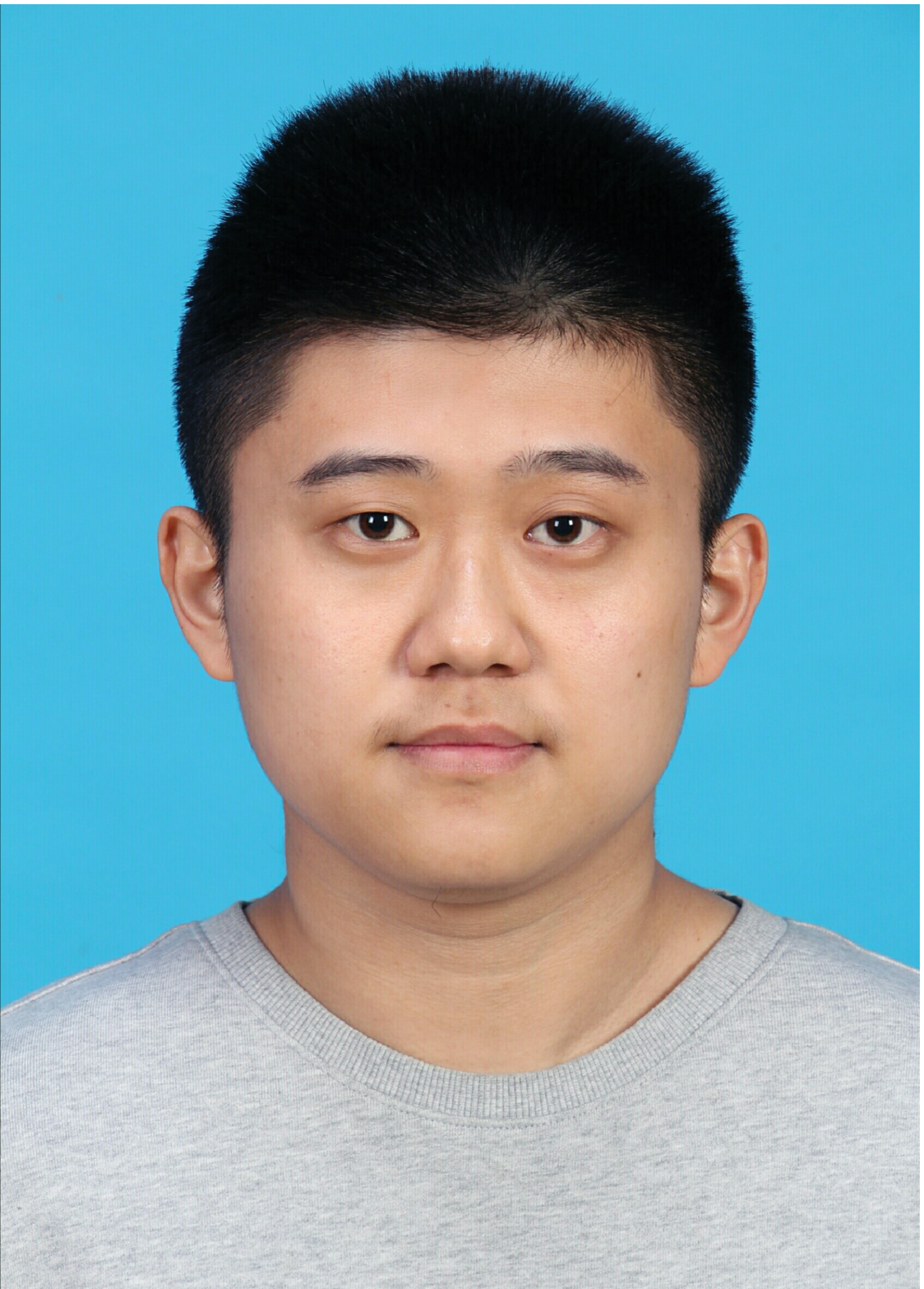}}]{Yuhang Jia}
received the B.S. degree in Qingdao University, China, in 2019. He is currently pursuing the master's degree with the Department of Electronic Engineering, Shanghai Jiao Tong University, China. His research interests include intelligent reflecting surface and convex optimization.
\end{IEEEbiography}
\begin{IEEEbiography}[{\includegraphics[height=1.25in]{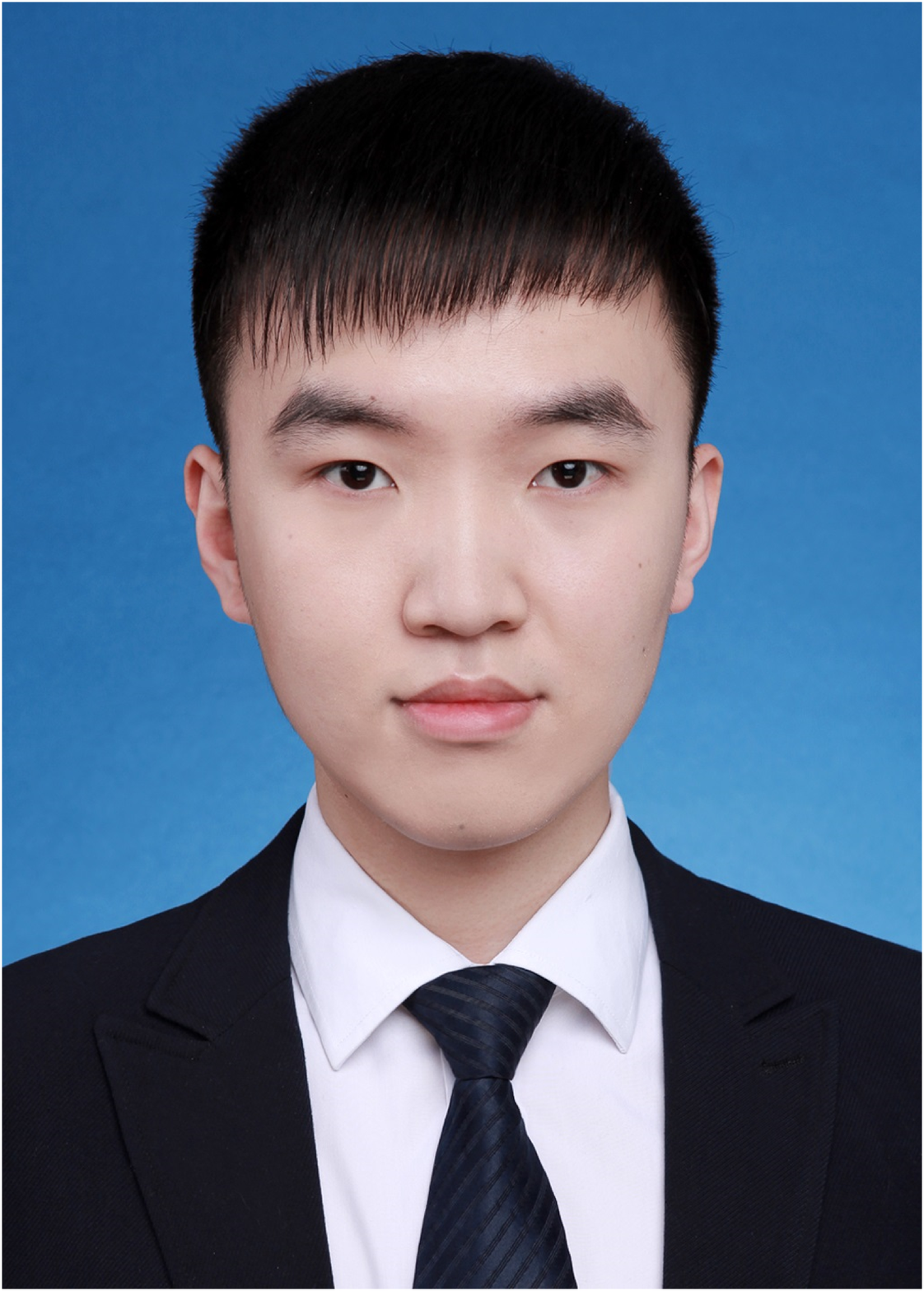}}]{Chencheng Ye}
received the B.S. degree in Shanghai Jiao Tong University, China, in 2018. He is currently pursuing the master's degree with the Department of Electronic Engineering, Shanghai Jiao Tong University, China. His research interests include cache-enabled wireless network, VR video transmission and convex optimization.
\end{IEEEbiography}

\begin{IEEEbiography}[{\includegraphics[height=1.25in]{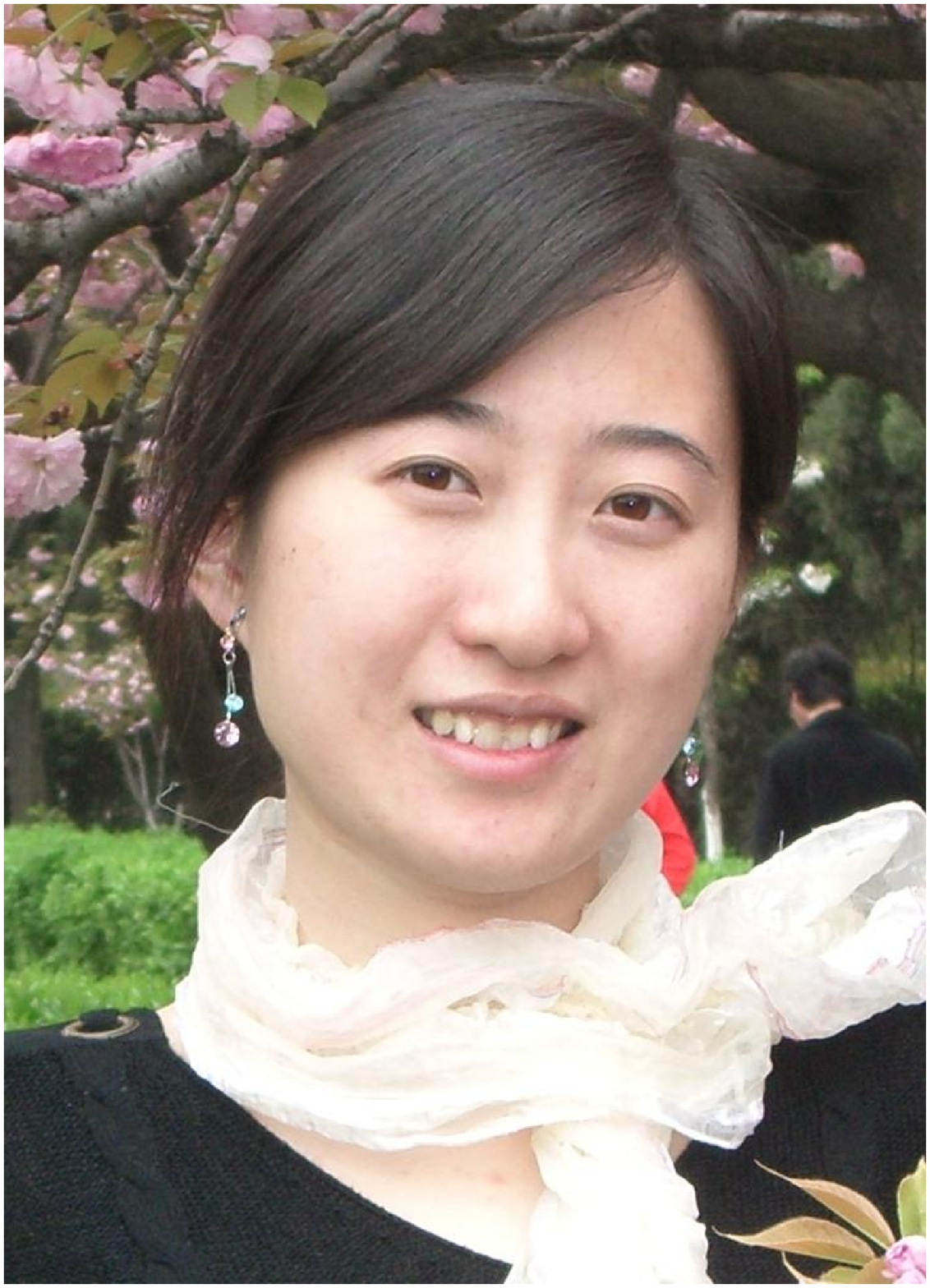}}]{Ying Cui}
received the B.E. degree in electronic and information engineering from Xi'an Jiao Tong University, China, in 2007, and the Ph.D. degree in electronic and computer engineering from the Hong Kong University of Science and Technology (HKUST), Hong Kong, in 2011. From 2012 to 2013, she was a Post-Doctoral Research Associate with the Department of Electrical and Computer Engineering, Northeastern University, Boston, MA, USA. From 2013 to 2014, she was a Post-Doctoral Research Associate with the Department of Electrical Engineering and Computer Science, Massachusetts Institute of Technology (MIT), Cambridge, MA. Since 2015, she has been an Associate Professor with the Department of Electronic Engineering, Shanghai Jiao Tong University, China. Her current research interests include optimization, cache-enabled wireless networks, mobile edge computing, and delay-sensitive cross-layer control. She was selected to the Thousand Talents Plan for Young Professionals of China in 2013. She was a recipient of the Best Paper Award at IEEE ICC, London, U.K., June 2015. She serves as an Editor for IEEE TRANSACTIONS ON WIRELESS COMMUNICATIONS.
\end{IEEEbiography}

\end{document}